\documentclass[generic,11pt]{imsart}

\usepackage{amsmath,amsthm,amssymb,color,multirow,algorithmic}
\RequirePackage[colorlinks,citecolor=blue,urlcolor=blue]{hyperref}

\usepackage[top=0.9in,bottom=1.1in,left=1.05in,right=1.05in]{geometry}
\usepackage[boxed]{algorithm}

\usepackage[numbers, sort&compress]{natbib}
\makeatletter 
\renewcommand\@biblabel[1]{#1.} 
\makeatother %

\usepackage{ifpdf}
\ifpdf
\usepackage[pdftex]{graphicx}
 \else
\usepackage[dvips]{graphicx}
 \fi

\startlocaldefs

\newcommand\E{\ensuremath{\mathbb{E}}}

\newcommand\cX{\ensuremath{\mathcal{X}}}

\newcommand\cL{\ensuremath{\mathcal{L}}}

\newcommand\cZ{\ensuremath{\mathcal{Z}}}
\newcommand\cE{\ensuremath{\mathcal{E}}}

\newcommand\cD{\ensuremath{\mathcal{D}}}

\newcommand\XX{\ensuremath{\mathbf{X}}}

\newcommand\R{\ensuremath{\mathbb{R}}}

\newcommand\F{\ensuremath{\mathcal{F}}}

\newcommand\PP{\ensuremath{\mathbb{P}}}
\newcommand\OO{\ensuremath{\mathcal{O}}}

\newcommand\eps{\epsilon}

\newtheorem{lemma}{Lemma}

\theoremstyle{remark}
\newtheorem{remark}{Remark}

\DeclareMathOperator\sign{sign}
\DeclareMathOperator\sgn{sign}
\endlocaldefs

\begin{document}



\begin{frontmatter}

\title{Sequential Design for Optimal Stopping Problems}
\runtitle{Sequential Design for Optimal Stopping Problems}

\author{\fnms{Robert B.} \snm{Gramacy}\ead[label=e2]{rbgramacy@chicagobooth.edu}}
\address{Booth School of Business\\ \printead{e2}}
\affiliation{The University of Chicago} \and
 \author{\fnms{Michael} \snm{Ludkovski}\ead[label=e1]{ludkovski@pstat.ucsb.edu}\thanksref{t1}}
 \thankstext{t1}{Partially supported by NSF ATD-1222262}
 \address{Dept of Statistics \& Applied Probability\\ \printead{e1}}
 \affiliation{University of California Santa Barbara}

\runauthor{Gramacy and Ludkovski}

\begin{abstract}
We propose a new approach to solve optimal stopping problems via simulation. Working within the backward dynamic programming/Snell
envelope framework, we augment the methodology of Longstaff-Schwartz that focuses on approximating the stopping strategy. Namely, we introduce adaptive generation of the stochastic grids anchoring the simulated sample paths of the underlying state process. This allows for active learning of the classifiers partitioning the state space into the continuation and stopping regions.
To this end, we examine sequential design schemes that adaptively place new design points
close to the stopping boundaries. We then  discuss dynamic regression algorithms that can implement such recursive estimation and local
refinement of the classifiers. The new algorithm is illustrated with a variety of numerical experiments, showing that an order of
magnitude savings in terms of design size can be achieved. We also compare with existing benchmarks in the context of  pricing
multi-dimensional Bermudan options.
\end{abstract}


\begin{keyword}
\kwd{optimal stopping} \kwd{regression Monte Carlo} \kwd{dynamic trees} \kwd{active learning} \kwd{expected improvement}
\end{keyword}

\end{frontmatter}

\section{Introduction}\label{sec:intro}
Numerical solution of optimal stopping problems remains a fertile area of research with applications in derivatives pricing, optimization of trading strategies, real options, and algorithmic trading.
As the underlying models continue to get more and more complex, the computational holy grail of robust, fast and accurate
solvers remains elusive. Essentially all analytic methods deteriorate in high-dimensional problems where geometric
intuition vanishes. Thus, recent attention has turned to probabilistic approaches, based on the Snell envelope representation. These
methods reduce to recursive estimation of conditional expectations
\begin{align}\label{eq:cond}
f_t(x)  := \E[ Y_{t+1} | X_t=x], \qquad t=T-1,T-2,\ldots, 0,
\end{align}
where the response is real-valued, $Y_{t+1} \in \R$, and the state process $X_t \in \R^d$ is a multi-dimensional Markov process,
typically with moderate dimension $d \in [1,10]$.

The estimation problem in \eqref{eq:cond} comes from a dynamic programming (DP) argument. Namely,
the process $(Y_t)$ is the Snell envelope of the payoff process $\{h_t(X_t)\}$, and its expected value $f_t(x)$ is interpreted as the price of the corresponding claim given initial condition $X_t =x$. The envelope $Y_{t+1}$ is defined recursively via
$
Y_{t+1} = h_{\tau_{t+1}} \left(X_{\tau_{t+1}} \right)$,
where $\tau_{t+1} := \inf\{ s \ge t+1 : h_s(X_s) \ge f_s(X_s) \} \wedge T$ depends on the future expectations $f_s(\cdot)$, $s > t$, cf.~\eqref{eq:first-hit}.

The seminal paper of Longstaff and Schwartz \cite{LS} proposed computing the Snell envelope by combining policy iteration dynamic
programming with a Monte Carlo approximation to \eqref{eq:cond}. The latter approximation used a cross-sectional regression over a Monte
Carlo sample of $X_t$, substituting pathwise continuation values as samples of $Y_{t+1}$. The key innovation over the earlier proposals in \cite{Carriere96,TsitsiklisVanRoy00} was the suggestion to only
use \eqref{eq:cond} for decision-making, rather than quantification of expected gains. Originally designed for the setup of American option pricing, the
algorithm of \cite{LS}, which we term RMC (Regression Monte Carlo), has become widely accepted in financial mathematics, insurance and
DP settings. It has also been implemented in many proprietary valuation systems employed by the financial industry. The great success of RMC is due to its flexible and simple implementation, as well as its strong empirical performance.

Nevertheless, much about RMC remains poorly understood and the algorithm must be fine-tuned for each use. These issues become of special
concern in high-dimensional, complex models. On the one hand,  a priori knowledge is usually unavailable in such cases so RMC-like
methods must run on ``auto-pilot". On the other hand, simulations become expensive and therefore computational space- and
time-constraints emerge. A variety of solutions, focusing specifically on the regression in RMC have been recently proposed, see
e.g.~\cite{Belomestny11,Kohler10review,Zanger13,TompaidisYang13}. At the same time, other simulation-based solutions beyond RMC continue to be developed, including quantization techniques \citep{PagesPrintems05}, Malliavin calculus methods
\cite{BouchardEkelandTouzi}, and stochastic mesh schemes \cite{BroadieGlasserman04,DelMoralHu10}.

In this paper, we propose an \emph{adaptive learning} version of RMC.  The key idea underpinning our approach is to view RMC as a sequential stochastic optimization problem. Indeed, in the context of optimal stopping the ultimate goal is not exact computation of the conditional expectation $f_t(\cdot)$ in \eqref{eq:cond} (which is identical to finding the
\emph{value function}), but the approximation of the zero level set of $f_t(\cdot)$, i.e., finding the boundary $\partial\{ x:
f_t(x) < 0 \}$. Indeed, the accuracy of RMC is entirely driven by a high-fidelity estimate of this \emph{stopping boundary} rather than
the full  map $x \mapsto f_t(x)$. Therefore, the statistical formulation of \eqref{eq:cond} is not about {regression} per se, but about contour-finding \cite{RanjanBingham08,PichenyGinsbourger13}.

To exploit this feature, we propose to use sequential design techniques to adaptively focus the computational efforts on classifying the sign of $f_t(\cdot)$ rather than on its global approximation. This optimization allows us to extract an order-of-magnitude savings in the simulation budget of RMC as measured by the corresponding sample sizes (though at the cost of increased regression/statistical modeling overhead).  Sequential design relies on auxiliary expected improvement (EI) metrics to preferentially sample at $x$-locations that maximize learning $\sign (f_t)$. This facilitates the construction of a spatially-adapted regression grid that efficiently approximates the stopping boundary. In order to handle such sequential grid design, the corresponding regression procedure must be updateable and localized, and moreover provide a measure of predictive uncertainty. Among this universe we suggest the use of dynamic trees \cite{GTP-trees11} which meet all these requirements and allow a straightforward implementation using publicly-available software.

The overall RMC scheme is then reduced to  a recursive sequence of contour-finding sub-problems, coupled with the usual DP framework. The iterative DP context introduces an extra dimension to budget allocation across time-steps and also a modified performance criterion.

The rest of the paper is organized as follows. In the remainder of Section \ref{sec:intro} we rigorously state the optimal stopping
problem we consider and the RMC solution framework. Section \ref{sec:lsmc} reviews existing implementations of RMC and serves as a
segue into Section \ref{sec:seq-rmc}, introducing our new sequential RMC methodology. Section \ref{sec:main} formalizes the new approach and describes the related sequential design schemes and dynamic tree regression modeling. Section \ref{sec:numeric}
presents several numerical examples, treating the classical problem of Bermudan option pricing  in Geometric Brownian motion and stochastic volatility models, and compares our algorithm with existing
benchmarks. Finally, Section \ref{sec:discuss} discusses potential for further improvement.

\subsection{Problem Statement}

We consider a discrete-time optimal stopping problem on a finite horizon. Let $\XX\equiv X_{1:T}$ be the state process, 
 a Markov chain on a stochastic basis $(\Omega, \mathcal{F}, \PP)$ taking values in  some (usually uncountable) subset $\mathcal{X} \subseteq \mathbb{R}^d$. The transition density
$p(t+1,X_{t+1}|t,X_t)$ of $\XX$ may not be available in closed form but can be easily simulated. A common example is when $\XX$ arises from a
nonlinear stochastic differential equation which is then discretized, e.g., via an Euler scheme.

\begin{remark}
Since we are interested in numerical schemes, we only consider the discrete-time setting. Any continuous-time problem can be resolved by
time-discretization with the usual estimates on the discretization error \cite{DelMoralHu10}. Similarly, infinite-horizon problems have
standard approximations using finite-horizon versions with $T$ large.
\end{remark}

Let $\F_t = \sigma( X_{1:t})$ be the information filtration generated by $\XX$,  $\mathcal{S} $ the collection of
all $\F$-stopping times smaller than some given horizon $T < \infty$, 
and $h_t : \mathcal{X} \to \mathbb{R}$ the reward function for stopping at $t=0,1,\ldots,T$. The optimal stopping problem consists of maximizing
the expected reward $h_\tau(X_\tau)$ over all $\tau \in \mathcal{S}$. More precisely, define for any $0 \le t \le T$,
\begin{align}\label{eq:Vt}
V(t,x) := \sup_{\tau \ge t, \tau \in \mathcal{S}} \E_{t,x} \left[ h_\tau(X_\tau) \right],
\end{align}
where $\E_{t,x}[\cdot] \equiv \E[ \cdot | X_t = x]$ denotes expectation given initial condition $x$.

Using the tower property of conditional expectations, we have that
\begin{align}\notag
V(t,x) &= \sup_{\tau \ge t, \tau \in \mathcal{S}} \E[ h_\tau(X_\tau) | X_t=x] \\ 
 & = \max \left( h_t(x), \E_{t,x}[ V(t+1, X_{t+1})] \right) = h_t(x) + \max( T(t,x), 0), \label{eq:recurse-V}
\end{align}
where we defined the timing-value $T(t,x)$ and continuation functions $C(t,x)$ via
\begin{align} \label{def:T} T(t,x) &:= C(t,x) - h_t(x); \\
\label{def:C}
C(t,x) &:= \E_{t,x} \left[ V(t+1,X_{t+1}) \right].
\end{align}
Since $V(t,x) \ge h_t(x)$ for all $t$ and $x$, an optimal stopping time $\tau^*(t,x)$ is characterized by
\begin{align}
\{ \tau^*(t,x) = t \} = \left\{ h_t(x) \ge C(t,x) \right\} = \{ T(t,x) \le 0 \}.
\end{align}
Thus, it is optimal to stop immediately if and only if the conditional expectation of tomorrow's reward-to-go is less than
the immediate reward. In other words, deciding whether it is optimal to stop at $t$
is equivalent to finding the zero level-set (or contour) of the function $T(t,x)$ or {classifying} the state
space $\mathcal{X} \ni X_t$ into the \emph{stopping region} $\mathfrak{S}_t := \{ x : T(t,x) \le 0 \}$ and its complement the continuation
region. We henceforth refer to $\mathfrak{S}_t$ as the {classifier} at time step $t$.
 By induction, the candidate optimal stopping time (if it exists) is
\begin{align}\label{eq:first-hit}
\tau^*(t,x) = \inf \{s \ge t : X_s \in \mathfrak{S}_s \} \wedge T.
\end{align}

\subsection{Monte Carlo Optimal Stopping}
Given any collection of classifiers $(\hat{\mathfrak{S}}_{1:T})$, one can approximate the
corresponding value of stopping by the empirical average payoff
 \begin{align}\label{eq:naive-mc}
  \hat{v}^{(N)}(t,x) \simeq \frac{1}{N}\sum_{n=1}^N h_{\tau^n} \left(x^{n}_{\tau^{n}} \right),
 \end{align}
where
$(x^{n}_{t:T})_{n=1}^N$ is a collection of $N$ independent simulated paths of the state process starting with $x^{n}_t=x$ and (cf.~Algorithm \ref{algo:forward-sim} below)
$\tau^{n} \equiv \hat{\tau}^{n}(t,x) := \inf\{ s \ge t: x^{n}_s \in \hat{\mathfrak{S}}_s \} \wedge T$, and $n=1,\ldots,N$.

Moreover, given the forward $\hat{\mathfrak{S}}_{t+1:T}$ one may use backward induction to construct $\hat{\mathfrak{S}}_t$ by estimating the respective timing value
$T(t,x)$ in \eqref{def:T} and its zero-contour. Indeed, a simulated path $x_{t:T}$ and corresponding pathwise stopping time
$\tau\equiv\tau(t+1,x)$ yields a realization $y_t := h_{\tau}( x_{\tau}) - h_t(x_t)$ of the random variable defining $T(t,x)$. In
particular, we have $\E[ y_t ] = T(t,x)$, or
\begin{align}\label{eq:cond-var}
Y_t(x) =  h_{\tau}( x_{\tau})-h_t(x)  = T(t,x) + \eps(t,x),
\end{align}
where the noise $\eps$ is mean-zero with state-dependent variance,
\begin{align*}
\E[ \eps^2(t,x)] \equiv \sigma^2(t,x) := \mathbb{V}ar \left( h_{\tau(t+1,x)}(X_{\tau(t+1,x)}) | X_t = x \right).
\end{align*}
Rather than providing a point estimate of $T(t,x)$, we are interested in the functional
 $x \mapsto T(t,x)$, and so
we view \eqref{eq:cond-var} as a regression problem, mapping inputs $x$ into random outputs $h_{\tau(t+1,x)}( X_{\tau(t+1,x)}) - h_t(x)$ that are centered around the true conditional expectation $T(t,x)$. Thus,  we generate $N$ forward paths by varying the sites $\{ x^n_t\}$, collect corresponding samples $\{y^{n}_t\}$, and then regress
 $\{ y^{1:N}_t\}$ on $\{x^{1:N}_t\}$ cross-sectionally to estimate $\hat{T}(t,x)$ and finally $\hat{\mathfrak{S}}_t$ from
\begin{align}\label{eq:estimate-St}
\hat{\mathfrak{S}}_t := \{ x: \hat{T}(t,x) < 0\}.
\end{align}

Overall, one can iterate backward over $t=T-1,T-2,\ldots, 0$ starting with the trivial $\mathfrak{S}_T = \cX$ to recursively add $\hat{\mathfrak{S}}_t$ to the collection $\hat{\mathfrak{S}}_{t+1:T}$. This approach directly approximates the stopping rule \eqref{eq:first-hit} while the value function can always be recovered from \eqref{eq:naive-mc}. Note that the right-hand-side in \eqref{eq:cond-var} depends on $\hat{\mathfrak{S}}_{t+1:T}$ and so we distinguish between the true Snell envelope $Y_t(x)$ based on the optimal $\tau^*(t,x)$, and the sampled $\tilde{Y}_t(x)$ that results from the recursive estimation and in turn determines $\hat{\mathfrak{S}}_t$. Algorithms \ref{algo:mc-generic} and \ref{algo:forward-sim} summarize this RMC method which allows a fully-simulation-based treatment of the Snell envelope.

Implementations of Algorithm \ref{algo:mc-generic} are distinguished in terms of how (i) the \emph{stochastic grids}
$\{ x^{1:N_t}_t\}$ are constructed, and how (ii) the classifier $\mathfrak{S}_t$ is estimated. The latter is a statistical modeling problem, while the former is a design problem (note that the $x^n_t$ are simulated, not specified a priori).
We use the shorthand $\cZ^{(N_t)}_t = \{x^{1:N_t}_t\}$ to refer to the size-$N_t$ design at step $t$ and, with a slight abuse of notation, also the respective pairs  $\{ (x_t, y_t)^{1:N_t}\}$ used for the regression.

\begin{algorithm}[H]
\caption{Regression Monte Carlo\label{algo:mc-generic}}
{\fontsize{11}{11}\selectfont\begin{algorithmic}[1]
\REQUIRE Design size $N_t$ for $t=T-1,\ldots,0$
\STATE $\hat{\mathfrak{S}}_T \leftarrow \cX $
\FOR[Step back in time]{$t=T-1,T-2,\ldots,1$} 
\STATE Generate the design $ \cZ^{(N_t)}_t := \{x^{n}_t\}$, $n=1,\ldots,N_t$
\STATE Using Algorithm \ref{algo:forward-sim} and $\hat{\mathfrak{S}}_{t+1:T}$ sample
$y^n_t$ (based on forward path $x^{(t),n}_{t+1:T}$) defined in \eqref{eq:cond-var} at each design point $x^n_t$
\STATE Regress $\{y_t^{1:N_t}\}$ against $\{x_t^{1:N_t}\}$ to estimate $\hat{T}(t,\cdot)$
\STATE Set $\hat{\mathfrak{S}}_t \leftarrow \{ x: \hat{T}(t,x) \le 0\}$
\ENDFOR
\STATE Starting with given initial condition $X^{n}_0 = X_0$ sample $v^n := h_{\tau^{(0),n}}( X^n_{\tau^{(0),n}})$, $n=1,\ldots,N$
\RETURN $\hat{v}^{(N)}(0,X_0) \simeq \frac{1}{N} \sum_{n=1}^N v^n$ as estimate of $V(0,X_0)$
\end{algorithmic}}
\end{algorithm}

\subsection{Least Squares Monte Carlo}\label{sec:lsmc}

The Longstaff-Schwartz \cite{LS} (henceforth LSMC) Algorithm is the original version of Algorithm \ref{algo:mc-generic} for optimal stopping. LSMC generates the designs $\cZ^{(N_t)}_t$ by a forward simulation of the $\XX$ state process starting with the fixed initial condition $X_0$,
$
x_{t+1}^{n} \sim p(t+1,\cdot| t,x^{n}_t)$,
where $p(t,x'|s,x)$ is the transition density of $\XX$.
Additionally, LSMC re-uses the same design points $\{x^n_s\}$ for the forward paths $x^{(t),n}_{t+1:T}$ used by Algorithm \ref{algo:forward-sim}, so that
 $x^{(t),n}_s = x^{(s),n}_s$ for any $s > t$. This minimizes simulation expenditures since only $\OO(T)$ total samples from the transition density of $\XX$ are required. The trade-off lies in introducing extra correlation between the approximations of $\mathfrak{S}_t$ for
 different dates, and the restriction to common design size $N_t \equiv N$ for all steps $t$.

Next, LSMC approximates the timing value $\hat{T}(t,\cdot)$ using a global least-squares regression. Recall that conditional expectation is {defined} as the
 $L^2$-minimizer, $C(t,x) = \arg\inf_{f(\cdot)} \|  V(t+1,X_{t+1})-f(X_t) \|_{L^2(\mathcal{F}_t)}$. LSMC approximates this
 infinite-dimensional minimization with a projection onto a finite-dimensional basis $span( B_1(x), \ldots, B_r(x))$ which
in practice amounts to parametric least-squares regression of
 $\{y_t^{1:N}\}$ against the basis functions $\{B_i(x^{1:N}_t)\}_{i=1}^r$.

 \begin{remark}
 Empirical performance of a least-squares estimator may be poor, especially if the distribution of the noise $\eps(t,x)$ in \eqref{eq:cond-var} is far from Gaussian. A variety of
solutions have been proposed to overcome this shortcoming, including use of kernel regression \cite{Belomestny11}, neural networks \cite{Kohler10nn}, radial basis
functions, smoothing splines and $L^1$-regularized regressions (all discussed in \cite{Kohler10review}). One particularly effective idea was suggested in Bouchard and Warin~\cite{BouchardWarin10} (henceforth BW11), relying on an adaptive partitioning strategy. Namely, one builds
a regular $k-d$ tree for a design $\cZ_t$ and then uses linear regression at each leaf. The partitioning is done recursively, splitting $\cZ_t$ into $N_p$ equi-probable cells along each dimension of $X_t$. Thus there are $(N_p)^d$ rectangularly
shaped cells in $\mathbb{R}^d$, containing an equal number of design points.  Spatial adaptivity is achieved, responding to the distribution of $X_t | X_0$ via a localized approach.   
\end{remark}

\begin{algorithm}[H]
\caption{Forward MC Algorithm for Optimal Stopping\label{algo:forward-sim}}
{\fontsize{11}{11}\selectfont\begin{algorithmic}[1]
\REQUIRE Classifiers $\mathfrak{S}_{t+1:T}$, initial point $x_t$
\STATE Sample $x_{t+1} \sim p(t+1,\cdot|t,x_t)$
\STATE $s \leftarrow t+1$
\WHILE{$x_s \notin \mathfrak{S}_{s}$ and $s < T$}
\STATE Sample $x_{s+1} \sim p(s+1,\cdot|s,x_s)$
\STATE $s \leftarrow s+1$
\ENDWHILE
\STATE $\tau \leftarrow s$
\RETURN payoff $y := h_\tau(x_\tau)$; trajectory $x_{t:\tau}$; stopping time $\tau$
\end{algorithmic}}
\end{algorithm}

\begin{remark}
LSMC reuses the grids $\{ x_s^{n}\}$, $s > t$ for obtaining $\{y^n_t\}$. This corresponds to
doing in-sample prediction for regression, creating an upward bias in estimated continuation values.  In the American Put option setting, this bias is observed to be significant (on the order of $> 1\%$ of final value), indicating a
 potential danger. In fact, most existing theoretical
 proofs \cite{EgloffKohlerTodorovic07,GlassermanYu,Kohler10nn,Zanger13} assume independent out-of-sample simulations $x^{(t),n}_{t+1:T}$ as stated in step 4 of Algorithm \ref{algo:mc-generic}. 
\end{remark}

\subsection{Loss Function}
Understanding the consistency of RMC requires determining whether the approximation $\hat{\mathfrak{S}}_t$ converges to the true
$\mathfrak{S}_t$. We note that Algorithm \ref{algo:mc-generic} is insensitive to errors in $\hat{T}(t,x)$ as long as it makes no impact on the estimated sign. This is dramatically different from value iteration methods based on \eqref{eq:recurse-V} where errors in $\hat{V}(t+1,\cdot)$ necessarily propagate back into $\hat{V}(t,x)$. Our basic tool is the following Lemma from Belomestny~\cite{Belomestny11}.

\begin{lemma}\cite[Lemma 5.1]{Belomestny11}
For any $t = 0,1,\ldots, T-1$, we have
\begin{align}
0 \le V(t,x) - \hat{V}(t,x) \le \E_{t,x}\left[ \sum_{s=t}^T |T(s,X_s)| \left(1_{ \{ \hat{\tau}=s, \tau^*=s\}} + 1_{\{\hat{\tau}>s, \tau^*=s\}} \right) \right]
\end{align}
where $\tau^*=\tau^*(s,X_s)$.
\end{lemma}

Using Cauchy-Schwarz inequality, it follows from the Lemma that
\begin{align}\label{eq:error-sets}
|V(0,x) -\hat{V}(0,x) | & \le \E_{0,x} \left[ \sum_{s=0}^T |T(s,X_s)| 1_{\{ X_s \in \cE_s \}} \right] \le \sum_{s=0}^T \| T(s,\cdot) \|^2_{L^2} \cdot \PP_{0,x}(\cE_s), \\
\text{where}\quad\cE_t &:= \mathfrak{S}_t \triangle \hat{\mathfrak{S}}_t = \{ x : \sgn T(t,x) \neq \sgn \hat{T}(t,x) \}
\end{align}
is the symmetric difference between the estimated and true stopping sets. Indeed on the event $\{ T(t,X_t) < 0 < \hat{T}(t,X_t) \}$ we stop too late, and on the event $\{ T(t,X_t) > 0 > \hat{T}(t,X_t) \}$ we stop too early. In both cases, the incurred loss in value is the (absolute value) of the true 
$T(t,X_t)$.

Recall that $\hat{T}(t,\cdot)$ is estimated based on the pathwise values $\tilde{Y}_t(x)$ in \eqref{eq:cond-var} which themselves depend on $\hat{\mathfrak{S}}_{t+1:T}$. Therefore, there is a double source of error in $\hat{\mathfrak{S}}_t$: from the use of the empirical design $\cZ_t$ and corresponding samples $y^n_t$, and from the sub-optimal payoffs recorded in some $y^n_t$'s due to propagation of error from $\hat{\mathfrak{S}}_{t+1:T}$. The usual strategy is to estimate these errors in terms of the $L^2$-norm $\| T(t,\cdot) - \hat{T}(t,\cdot) \|_{L^2(X_t)}$ via (cf.~\cite[Prop.~6.1 and (6.25)]{Egloff05})
\begin{align}\label{eq:egloff}
\| T(t,\cdot) - \hat{T}(t,\cdot) \|_{L^2(X_t)} \le Err_t + 3 \| \tilde{Y}_t(X_t) - Y_t(X_t) \|_{L^2(X_t)},
\end{align}
where $Err_t = Err_t(\cZ_t)$ is the estimation error for $\mathfrak{S}_t$ based on a design $\cZ_t$. The back-propagation term above can be controlled by (\cite[Prop.~6.4]{Egloff05})
$$
\| \tilde{Y}_t(X_t) - Y_t(X_t) \|_{L^2(X_t)} \le \sum_{s=t+1}^T \| \hat{T}(s,\cdot) - T(s,\cdot) \|_{L^2(X_s)},
$$
leading to a Gronwall-type estimate for the total RMC error in terms of $Err_t$. Assuming that there exists $K$ such that $\|\tilde{Y}-Y\|_{\infty} \le 2K$ is bounded (which follows as soon as the payoffs $\|h_t\|_\infty \le K$ are bounded, achievable via truncation if necessary), we also have the more basic estimate
\begin{align*}
\| \tilde{Y}_t(X_t) - Y_t(X_t) \|^2_{L^2(X_t)} & \le (2K)^2 \PP_{0,x}( \tilde{Y}_t(X_t) \neq Y_t(X_t)) \\
& \le (2K)^2 (1-\PP_{0,x}( X_s \notin \cE_s \forall s \ge t+1)) \le (2K)^2 \sum_{s = t+1}^T \PP_{0,x}( \cE_s),
\end{align*}
which shows the direct influence of the error sets $\cE_t$ on \eqref{eq:egloff}.

The local error $Err_t(\cZ_t)$ in \eqref{eq:egloff} depends on the quality of the statistical model or architecture used to produce $\hat{T}(t,\cdot)$ and the size (and quality) of the design $\cZ_t$. The LSMC recipe of parametric least-squares regression offers a convenient theoretical framework for this analysis. It expresses $Err_t$ in terms of the design size $|\cZ_t|=N$ and the  best possible approximation error $\inf_{f \in \mathcal{H}_N} \| T(t,\cdot) - f \|_2$ where the function class $\mathcal{H}_N$ depends on $N$, and allows consideration of the convergence speed in the limit $N \to \infty$. See the original functional central limit result in \cite{ClementLP} and subsequent generalizations in \cite{Egloff05,EgloffKohlerTodorovic07,Zanger13}. It remains an open problem to express $Err_t$ directly in terms of $\PP_{0,x}(\cE_t)$, though see some results in this direction in \cite{Belomestny11}.

Returning to \eqref{eq:error-sets}, we now remark that the loss function $\cL(\hat{T}; T)$ for judging the accuracy of $\hat{T}(t,\cdot)$ is in fact rather different from the least-squares criterion. Indeed, rather than minimizing the $L^2$-norm $\| \hat{T}(t,\cdot) - T(t,\cdot)\|_{L^2(X_t)}$ that is usually considered, we really ought to control the approximation of $\mathfrak{S}_t$ which is equivalent to minimizing the pointwise loss
\begin{align}\label{eq:loss}
L^{(zc)}(\hat{y},y) =  |y| \cdot I \left(\sgn y \neq \sgn \hat{y} \right)
\end{align}
between the true $y$ and its estimator $\hat{y}$. Re-expressing \eqref{eq:error-sets} in terms of $L^{(zc)}$ yields that the global approximation error at step $t$ is
\begin{align}\label{eq:total-error}
\cL(\hat{T}(t,\cdot); T(t,\cdot)) = \hat{\E}_{0,x} \left[ L^{ZC}(\hat{T}(t,X_t), T(t,X_t)) \right],
\end{align}
i.e.~the expected loss averaged using the law  $\hat{\PP}_{0,x}$ which defines the dynamics of the \emph{controlled} $(X_t)$ starting at the fixed initial condition and stopped at $\hat{\tau}$.
Thus, regions of $\cX$ that are likely to be reached by the controlled $X_t$ are naturally given more influence for expected loss. Because $\hat{\PP}_{0,x}$ depends on the stopping rules on the entire $\{0,1,\ldots,T\}$, it is not directly available at intermediate $t$ and will be naturally approximated by the original $\PP_{0,x}$.

\section{Sequential RMC}\label{sec:seq-rmc}
The loss criterion in \eqref{eq:total-error} is localized, especially compared to the global $L^2$ criterion. As such, the accuracy of the algorithm is crucially dependent on the stochastic grids. This idea underlies our proposal to extract new computational efficiency by optimizing these grids. Formally, given a computational budget of $N$ locations, we wish to \emph{design} the collection $\{ x^{1:N}_t\}$ so as to minimize the resulting expected loss from the statistical model of $\mathfrak{S}_t$.

\subsection{Guide to the Algorithm}
In the next Section we describe such an approach that achieves this goal by {iteratively} augmenting the designs. As our numerical experiments demonstrate, this allows roughly an order of magnitude savings in the grid size.
Before detailing the proposed implementation, which requires review of concepts from statistical decision theory [Section \ref{sec:rsm}], we provide an intuitive overview of the sequential version of Algorithm~\ref{algo:mc-generic}.

Roughly speaking the quality of the approximation to $T(t,\cdot)$ is controlled by the local \emph{density} of design points. Hence, a good approximation of its zero-contour requires placing the grid points in the vicinity of the stopping boundary $\partial \mathfrak{S}_t$. 
Since the structure of $\partial \mathfrak{S}_t$ is unknown a priori, we adaptively build the designs to ``zoom in'' towards the estimated $\partial \mathfrak{S}_t$. To do so, instead of fixing  $\{ x^{1:N_t}_t\}$, we generate a sequence of increasing designs $ \cZ^{(n)}$, $\cZ^{(n)} \subseteq \cZ^{(n+1)}$, $n=1,\ldots, N_t$. Thus, step 3 in Algorithm \ref{algo:mc-generic} is replaced with a loop.

\begin{remark}
To guide the grids $\cZ$ towards the stopping boundary, one could borrow ideas from stochastic simulation, such as importance
sampling. The main difficulty with using importance sampling is the need to have \emph{a priori} information about $\mathfrak{S}_t$. We
refer to \cite{Moreni04,DelMoralHu12} for details on combining importance sampling and optimal stopping.
\end{remark}

Each design $\cZ^{(n)}$ induces an estimate $f^{(n)}$ of $T(t,\cdot)$.
Given an existing $\cZ^{(n)}$, the design sub-problem is to optimally augment it with a new design point $x^{n+1}_t$ and corresponding realization $y^{n+1}_t$. The value of $(x^{n+1}_t, y^{n+1}_t)$ is judged in terms of the over-arching objective of maximizing approximation accuracy based on $\cL$ from \eqref{eq:total-error}. Indeed, the new $\cZ^{(n+1)}$ will induce the corresponding fit $f^{(n+1)}(\cdot)$, which in turn is evaluated through a loss functional $\tilde{\cL}(f^{(n+1)})$ to be defined.
Hence, we take $x^{n+1}_t$ to be a minimizer of the expected loss $\E[ \tilde{\cL}( f^{(n+1)}) | \cZ^{(n)}]$. The latter expectation integrates over the distribution of $Y_t(x^{n+1}_t)$. This approach to designing $\cZ^{(N)}$ can be viewed as a myopic greedy procedure since it focuses on the one-step improvements in $\cL$. To guard against over-optimism, it must also guarantee that $\cL( f^{(n)}; f) \to 0$ as $n \to \infty$.

The loss functional $\tilde{\cL}( \hat{f})$ is an empirical approximation of the true loss $\cL( \hat{f}, f)$ which cannot be evaluated because $f$ is unknown.
This is achieved by viewing the true $T(t,\cdot)$ as a random object, so that given $\cZ^{(n)}$ we consider the pointwise  posterior distribution
\begin{align}
T(t,x) | \cZ^{(n)}\sim F_x( \cdot | \cZ^{(n)}).
\end{align}
For example, a typical estimator $\hat{f}(x) = \E[ T(t,x) | \cZ^{(n)}]$ is the pointwise posterior mean. Given the estimated $\hat{F}_x$ and $\hat{f}$ we can easily obtain an estimate of \eqref{eq:loss}
\begin{align}
\tilde{L}(\hat{f})(x) = \int_\R L^{(zc)}(\hat{f}(x), y) \hat{F}_x( dy |  \cZ^{(n)}),
\end{align}
i.e.~the local loss averaged with respect to the posterior of $T(t,x)$,
and finally the global error
\begin{align}\label{eq:empirical-error}
\tilde{\cL}(\hat{f}) := \int_\cX \tilde{L}(\hat{f})(x) p(t,dx|0,X_0).
\end{align}
The latter object is the empirical analogue of the stepwise loss $Err_t$ in \eqref{eq:egloff} and hence permits adaptive control of the associated error propagation.

\begin{remark}
$\tilde{\cL}$ is optimistic about the true error $Err_t$ since it ignores the error-in-variables propagation, i.e.~that some of the sampled $Y$'s are not correct. See \cite{KohlerFromkorth13} for the related error analysis.
\end{remark}

\subsection{Computational Methodology}

Implementing the sequential design procedure is challenging from two directions.
First, the loss function $\tilde{\cL}$ in \eqref{eq:empirical-error} evaluates the global accuracy of $f^{(n)}$ and is intuitively similar to the integrated (or expected) mean-squared regression error over the input space. Consequently, evaluating $\cL(f^{(n+1)})$ requires understanding the impact of adding a new design point on the regression fit, which is generally intractable. Instead, we therefore rely on a local expected improvement procedure, which provides surrogate scores $EI^{(n)}(x)$ for any $x \in \cX$ and is easy to evaluate. Roughly speaking, we thus replace the expected \emph{global} improvement criterion with a local one. Second, finding the minimizer $x^{n+1,*}$ leads to an optimization problem over $\cX$. For continuous multi-dimensional state spaces $\cX$, this is a costly task. In fact, since this is just an intermediate step in the overall RMC method, the requisite accuracy is not so important. Thus, we propose simple approximate procedures to pick $x^{n+1}$. In particular, we select $x^{n+1}$ probabilistically, which provides extra guarantees on the convergence of the method and is computationally quicker.

The sequential design approach replaces the single fitting step 5 in Algorithm \ref{algo:mc-generic} with a sequential regression loop. The corresponding regression method must be adapted to this setting, in particular it ought to be \emph{updateable} and provide good estimates of the posterior distribution $F_x$. Furthermore, the modeling must be localized in order to handle increasing density of design points around $\partial \mathfrak{S}$. As a result, standard linear parametric models (which are updateable and fast) are not appropriate since adding design points induces \emph{global} changes in $\hat{f}$ and hence little control on the expected $\tilde{\cL}(f^{(n+1)})$. Instead, we propose dynamic trees (DTs) \cite{GTP-trees11} which is a non-parametric Bayesian tree-based regression method. DTs use a divide-and-conquer approach to efficiently refine the fits as the design zooms in towards $\partial \mathfrak{S}$. Moreover, DTs provide conditionally Gaussian posterior distributions for $F_x$ and hence are convenient to use
with expected improvement rules.

We stress that our specific choices of sequential design and regression methods are not meant to be authoritative. Indeed, there are many other possible combinations for how to augment $\cZ^{(n)}$ and build $f^{(n)}$, see discussion in Sections \ref{sec:al} and \ref{sec:dyna-trees} below. Thus, the numerical results herein are primarily illustrative and we expect that ultimately even more efficient implementations (possibly with further customization) will be found.

\subsection{Illustration}
\begin{figure}[ht!]
\begin{centering}
\begin{tabular}{ccc}
\begin{minipage}{0.32\textwidth}
\includegraphics[width=0.97\textwidth]{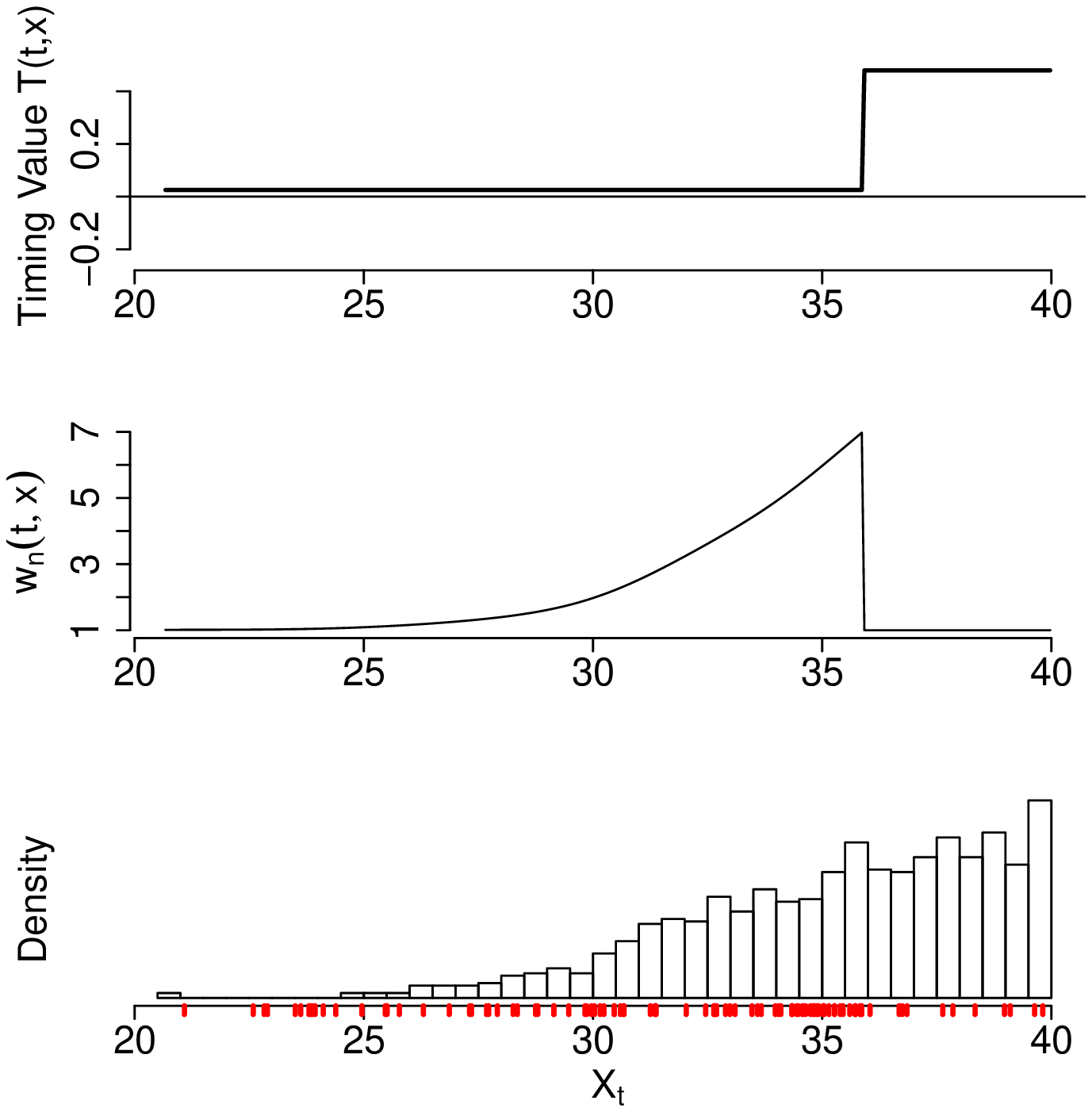} 
\end{minipage} &
\begin{minipage}{0.32\textwidth}
\includegraphics[width=0.97\textwidth]{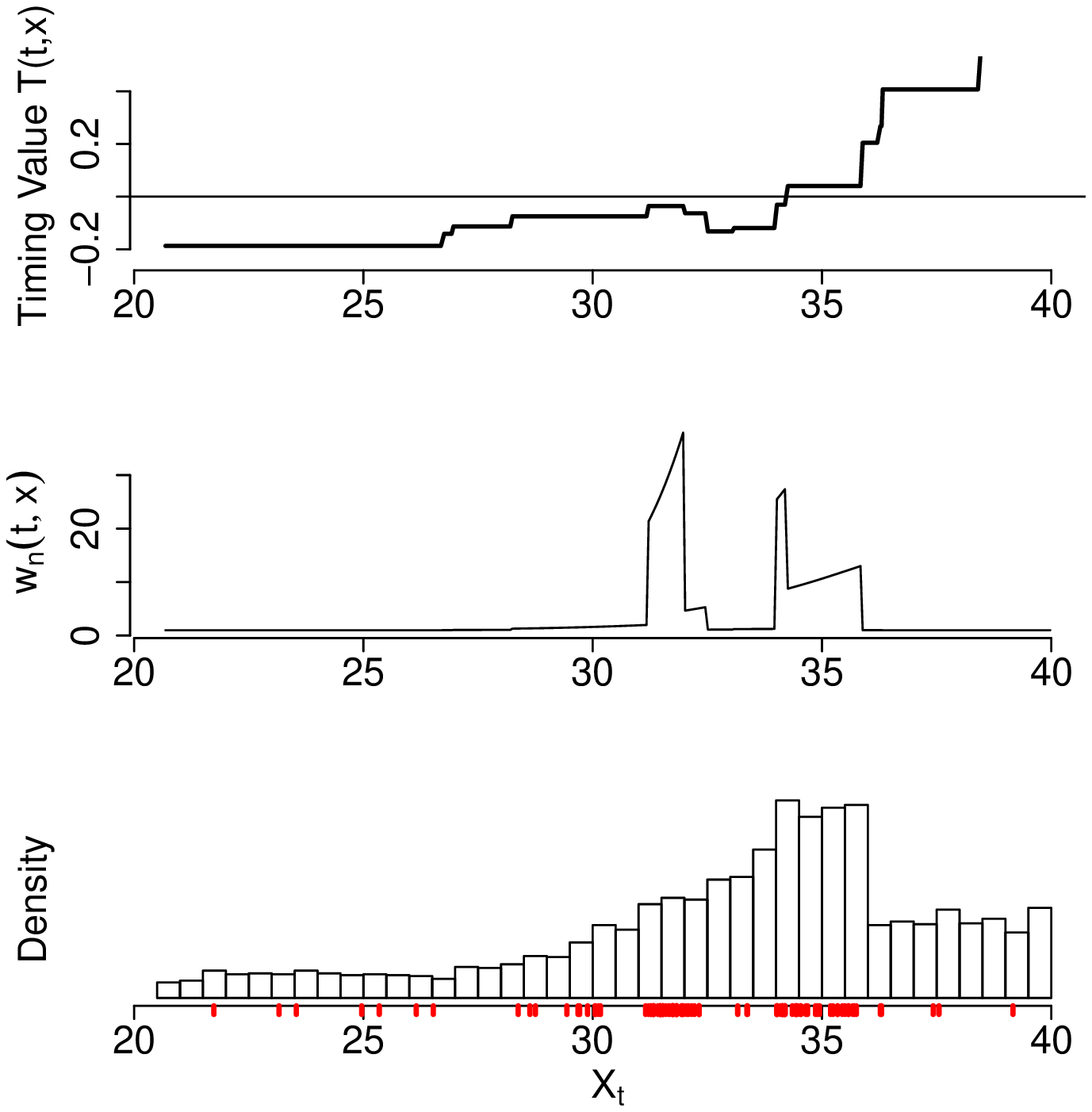}
\end{minipage} &
\begin{minipage}{0.32\textwidth}
\includegraphics[width=0.97\textwidth]{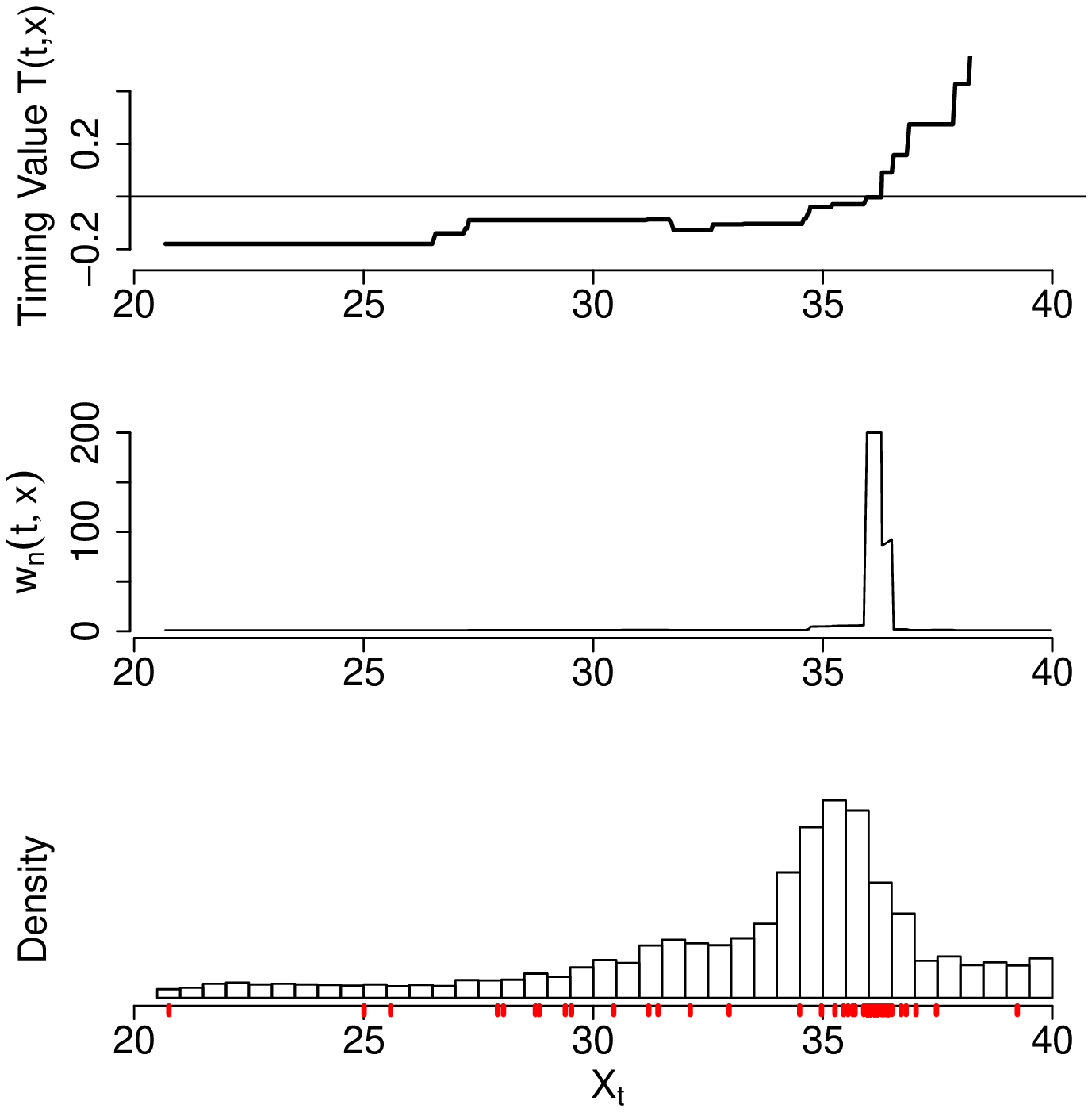}
\end{minipage}  \\
$n=1000$ & $n=3000$ & $n=5000$
\end{tabular}
\begin{minipage}{0.92\textwidth}
\caption{ \label{fig:active-learning} Active learning of the classifier $\mathfrak{S}_t$. We begin with an initial design
$\{x^{1:N_0}_t\}$ of $N_0=1000$ points sampled from $p(t,X_t |0,X_0)$, after which $4,000$ more design points are added, with $N'=100$
samples between recomputing $f^{(n)}$. The final design size is thus $N_t = 5,000$. The top panels show fits using $M=4$-particle  dynamic trees with constant leafs. Middle panels show the
resulting $x \mapsto w_n(t,x)$ weights (cf.~\eqref{eq:weights}) used in sequential design. Bottom panels show the histograms of intermediate designs $\cZ^{(n)}$ and the locations of the latest samples $\{x^{n:n+N'}_t\}$ (red hashmarks). Full details are in Section~\ref{sec:example-1d}.}
\end{minipage}
\end{centering}
\end{figure}

Figures \ref{fig:active-learning} and \ref{fig:compBW-1d} illustrate the new RMC paradigm for the classical example of a 1-dimensional Bermudan Put option in a Black-Scholes model. Figure \ref{fig:active-learning} shows the sequential design phase. As explained, we begin with a small design $\cZ^{(N_0)}$ and gradually augment it as realizations $y^{n}_t$ are collected. At each step, the fits $f^{(n)}$ (top panel of Figure \ref{fig:active-learning}) are updated and the candidate locations for $x^{n+1}_t$ are ranked in terms of an expected improvement function $EI_n(x)$ (middle panel of Figure \ref{fig:active-learning}). As a result, the grids $\cZ^{(n)}$ increasingly concentrate around $\partial \hat{\mathfrak{S}}_t$ (bottom panel of Figure \ref{fig:active-learning}). In this 1-d example, the true stopping boundary $\partial {\mathfrak{S}}_t$ is known to be a single point $\underline{s}_t$.

By using specialized regression methods and adaptive design, the final output of the RMC at time $t$ substantially differs from previous approaches. Figure \ref{fig:compBW-1d} illustrates the new features of our approach. Compared to the benchmark (based on the implementation in \cite{BouchardWarin10}), the adaptive $\cZ^{(N)}$ is sharply focused on identifying the stopping boundary, which in turn permits a much higher degree of refinement in estimating $\hat{T}(t,x)$ for $x$ in the neighborhood of $\underline{s}_t$. As a result, comparable accuracy is achieved for much smaller design sizes (in the Figure a simple eye-test reveals that our method beats out the benchmark despite using a design eight times smaller). Intuitively, in the non-adaptive method more than 80\% of the simulations are entirely wasted (or even harmful since they may actually degrade the approximation quality). Our approach minimizes this inefficiency.

\section{Sequential Design for Optimal Stopping}\label{sec:main}

\subsection{Posterior Fit Uncertainty}\label{sec:rsm}
We summarize the response surface methodology as applied to RMC. Below the reader should substitute $f(x) =T(t,x)$.

Consider an unknown continuous Lipschitz response function $x \mapsto f(x)$, treated as a random element of the model space $C_{Lip}(\cX)$ to be learned. The function $f$ can be noisily sampled via
\begin{align}\label{eq:regress}
 Y = f(X) + \eps, \qquad \E[ \eps] =0, \quad \mathbb{V}ar(\eps |X) = \sigma^2(X).
\end{align}
Thus, the observed response $Y(x)$ is an unbiased sample of $f(x)$ with observation variance $\sigma^2(x)$. We assume that we do not know either $f(x)$ or $\sigma^2(x)$ but have access to a simulation engine that for any site $x \in \cX$  can generate independent samples $(x,Y(x))$ from \eqref{eq:regress}.

Consider existing sampled data $\cZ^{(n)} := (x,y)^{1:n}$. Based on this sample we aim to construct a posterior distribution of $f$, namely
\begin{align}
f(x) | \cZ^{(n)} \sim F_x( \cdot; \cZ^{(n)}).
\end{align}
The global posterior surface $F$ is a measure on $C_{Lip}(\cX)$; we primarily
focus on pointwise posteriors $F_x$ which are measures on $\cX$. In
particular, we distinguish the posterior mean $m^{(n)}(x)$ and variance
$v^{(n)}(x)$ surfaces, where the superscripts remind us that these depend on
$(x,y)^{1:n}$.

Given a loss function, for any location $x \in \cX$, some summary statistic from $F_x( \cdot | \cZ^{(n)})$ can then be used to form the estimator $\hat{f}^{(n)}(x)$ of $f(x)$. For sequential RMC, our estimator should be adapted to the pointwise loss in \eqref{eq:loss}. For simplicity we rely throughout on the posterior mean $\hat{f}^{(n)}(x) \equiv m^{(n)}(x)$; see \cite{GL14stat} for further discussion, including consideration of other metrics such as the posterior median.

The precise representation of the posterior $F_x(\cdot)$ is regression-method specific. Two convenient setups are ensemble methods and Gaussian posteriors. Assuming $n$ large, a central limit argument suggests that $F_x$ should be approximately normal, i.e.
\begin{align}\label{eq:normal-pred}
f(x) | \cZ^{(n)}  \sim F^{Gsn}_x \equiv \mathcal{N} \left( m^{(n)}(x),  v^{(n)}(x) \right).
\end{align}

Ensemble methods follow the opposite rule and build a fully empirical posterior $F_x$. Namely, the regression model consists of a collection of $M$ point estimates $\hat{f}_m(x)$ which are combined together through weighted averaging to obtain $\hat{f}(x) = \sum_{m=1}^M w_m(x) \hat{f}_m(x)$ and the discrete posterior
\begin{align}\label{eq:discrete-F}
F^{disc}_x(\cdot) = \sum_{m=1}^M w_m(x) \delta_{\hat{f}_m(x)}(\cdot),
\end{align}
where $\delta_z(\cdot)$ is the Dirac delta measure and $w_m$ (typically $w_m = 1/M$) is the weight of the $m$-th estimator.
Such ensemble methods are competitive in multi-dimensional nonlinear settings and include random forests, generalized boosted models, bagged trees and particle-based models (such as dynamic trees and particle learning Gaussian processes).
Our specific way of modeling the posterior $F_x$ uses both \eqref{eq:normal-pred} and \eqref{eq:discrete-F}
and  is described in Section \ref{sec:dyna-trees}.

\subsection{Expected Improvement}\label{sec:al}
Intuitively, the aim of sequential design is to sequentially explore the input space $\mathcal{X}$ towards placing samples on the (unknown) $\partial \mathfrak{S}_t$, so as to maximize its estimation accuracy. The main trade-off is between sampling close to the \emph{current
estimate} of the contour versus reducing the predictive uncertainty of the fit. Note that refining the estimated boundary $\partial \hat{\mathfrak{S}}_t$ is relative to the regions of $\cX$ most likely to be visited by $X_t$, and must be done such that $v(x) \to 0$ as $n \to \infty$ in order to guarantee global consistency. The basic strategy is then to construct a heuristic,
$EI_n(x)$, which measures the information gain at a location $x \in \cX$ conditional on existing samples
$(x, y)^{1:n}$ and is used to guide the selection of the next design point $x^{n+1}$.

We use the term \emph{expected improvement} (EI), originating in \cite{JonesSchonlauWelch98}, to refer to
these scores.  
Other sequential design approaches include active learning \citep{Mackay92,cohn:1996} and stepwise uncertainty reduction \cite{Picheny12}.
See \cite{GramacyPolson11} for discussion of EI/AL alternatives for regression and classification.
Generally, EI heuristics are tied to the posterior uncertainty of the fit and rely on the normal approximation \eqref{eq:normal-pred}.
Hence, the different EI scores are specified as functions of $m^{(n)}(x)$ and $v^{(n)}(x)$.

To construct our EI score we merge ones designed to identify the contour and reduce posterior uncertainty. We search for contours via the empirical loss at a location $x$ using \eqref{eq:loss} and \eqref{eq:normal-pred}:
\begin{align}\notag
L^{(zc)}_n(x) = \tilde{L}^{(zc)}(m^{(n)})(x) & = \int_\R |y| 1_{\{\sgn y \neq \sgn m^{(n)}(x) \}} F^{Gsn}_x(dy) \\ \label{eq:EI-zc}
 & = \sqrt{v^{(n)}(x)} \,\phi \left( \frac{-|m^{(n)}(x)|}{\sqrt{v^{(n)}(x)} } \right) - |m^{(n)}(x)| \Phi\left(  \frac{-|m^{(n)}(x)|}{\sqrt{v^{(n)}(x)} } \right),
\end{align}
 where $\phi(\cdot), \Phi(\cdot)$ are the standard Gaussian density and cdf respectively. A simpler version is based on the posterior probability of correctly estimating the sign of $f(x)$:
\begin{align}\label{eq:EI-sign}
L^{(sgn)}_n(x) & = \int_\R 1_{\{\sgn y \neq \sgn m^{(n)}(x) \}} F^{Gsn}_x(dy) 
  = \Phi \left( -\frac{|m^{(n)}(x)|}{\sqrt{v^{(n)}(x)} } \right).
\end{align}
We then consider the expected reduction in posterior variance based on the {\em active learning Cohn} (ALC) criterion \cite{cohn:1996}.
\begin{align}
ALC_n(x) := \E[ v^{(n+1)}(x) - v^{(n)}(x) | x^{n+1}=x]
\end{align}
The expectation on the right-hand-side can be evaluated explicitly with our regression choice, see \eqref{eq:DT-alc}. Note that $ALC_n(x)$ tends to be large in regions where either (i) density of the design $\cZ^{(n)}$ is relatively low; or (ii) the variance of the noise $\eps(x)$ is high. Finally,
recall that the overall criterion \eqref{eq:total-error} also takes into account the underlying distribution $p(t,\cdot | 0, X_0)$ of $X_t$. We propose to combine the above three components into a single EI score via
\begin{align}\label{eq:ei-main}
EI_n(t,x) := L_n(x) \cdot ALC_n(x) \cdot p(t,x|0,X_0).
\end{align}
The above choice prefers locations close to the relevant contour (high $L_n(x) \cdot p(t,x|0,X_0)$) but also guards against myopia by taking $ALC_n(x)$ into account. Without the latter adjustment, all the points would be placed right around the estimated $\partial \mathfrak{S}$ resulting in insufficient exploration.

\begin{remark}
There are other possibilities of blending together $L(x)$ and $ALC(x)$. For example, the UCB-type criteria from sequential learning suggest
\begin{align}\label{eq:EI-ucb}
EI^{(ucb)}_n(x) & = -| m^{(n)}(x)| + \gamma_n ALC_n(x).
\end{align}
The parameter $\gamma_n$ is user-defined and balances the dual preference for design points close to the contour (where $m^{(n)}(x) \simeq 0$) and where posterior variance of $f| \cZ^{(n+1)}$ can be most reduced. From our numerical experiments we found that tuning $\gamma$ is difficult, not least because different $\gamma$ seem to be needed for different time steps $t$ since the signal-to-noise ratio is time dependent. Also, it is not clear how to combine a UCB criterion with the underlying weights $p_{X_t|X_0}(x)$. Our proposed criterion \eqref{eq:ei-main}
appears to be rather robust when combined with probabilistic sampling.
\end{remark}

To guarantee consistency as $n\to\infty$, the sequential design rule must ensure that $\lim \cL(\hat{f}^{(n)}) \to 0$. As such, the ultimate rule for picking $x^{n+1}$ must guarantee that the density of design points grows without bound on the full $\cX$.
We address these concerns in Section \ref{sec:sampling}, in particular via probabilistic sampling of $x^{n+1}$.

\begin{remark}
The sequential design iterations are analogous to stochastic optimization which aims to find the global maximizer $x^*$ of some response $x \mapsto g(x)$ given a sequence of noisy measurements $(x, y)^{1:n}$. In our context, the main modification is that contour-finding seeks not a finite set of local maxima, but a whole $(d-1)$-dimensional hyperplane $\{ x : T(t,x) = 0 \}$. Moreover, in contrast to classical settings where a unique solution is assumed to exist, the geometry of the boundary $\partial \mathfrak{S}_t$ (e.g.~the number of zero-crossing points in one-dimension) is unknown.
\end{remark}

\subsection{Dynamic Trees}\label{sec:dyna-trees}
To implement sequential RMC we propose the framework of Dynamic Trees \cite{GTP-trees11}. DTs offer a thrifty sequential nonparametric non-linear regression with conditionally Gaussian predictive equations. This is achieved via a fully
Bayesian representation of the response surface which combines simple fits at the leafs (constant or linear models) with a flexible
multiple-tree representation of the global fit.  As such, dynamic trees are well-suited to our context by allowing explicit calculation of losses \eqref{eq:EI-zc}-\eqref{eq:EI-sign} for new locations, and easy updating of the fits $f^{(k)}$'s that organically grow  to refine local estimates as sample density increases.

Trees, generically, are a classic nonlinear and nonparametric regression or classification model. They recursively partition the
multidimensional input space $\mathcal{X}$ into a number of hyper-rectangles such that nearby inputs with similar output $Y$ values fall
within the same hyper-rectangle. This partitioning scheme gives rise to a set of if-then-else rules that can be represented graphically
as a tree. Bayesian regression trees \cite{chip:geor:mccu:1998} are specified by a prior distribution on how the input space can be
recursively partitioned and a likelihood comprising a product of simple  regression models applied independently in
each partition.  Dynamic trees specify a similar process for how trees evolve as new data arrive, which makes them particularly well suited to streaming
data.  At iteration $k$, after having seen data $(x,y)^{1:k}$ and inferred a tree $\mathcal{T}_k|(x,y)^{1:k}$, a simple set of
stochastic rules defines which $\mathcal{T}_{k+1}$ may be considered when $(x^{k+1}, y^{k+1})$ arrives. In this process, the new
$\mathcal{T}_{k+1}$ must be identical to the old $\mathcal{T}_k$ except near the leaf node $\eta(x^{k+1})$ containing $x^{k+1}$.  The
process stochastically chooses from three local modifications based on support from $y^{k+1}$ in the posterior distribution: {\it  keep}
$\eta(x^{k+1})$ unchanged in $\mathcal{T}_{k+1}$; {\it grow} a new split, making $\eta(x^{k+1})$ a parent of two new leafs in
$\mathcal{T}_{k+1}$; or {\it prune} the tree to make the parent of $\eta(x^{k+1})$ a leaf in $\mathcal{T}_{k+1}$.  A particle
approach---essentially applying these rules independently to $M$ similar trees grown stochastically on the same data---can reduce Monte
Carlo error (via averaging) and lead to more accurate uncertainty quantification (by studying the spread of trees).  Finally, dynamic
trees apply a sequential Monte Carlo, or ``filtering,'' approach  that appropriately couples the particles/trees via particle resampling mechanisms to offer further statistical efficiency gains. The {\sf R} package {\tt dynaTree} \cite{GTP-trees11} implements the above methodology and was used for our examples below.

Because dynamic trees are an instance of a local averaging forecaster \cite{Gyorfi}, global consistency follows as long as tree sizes grow without bound and the number of design points in each leaf grows sufficiently quickly as a function of leaf depth in the tree. Both of the above can be obtained with a suitable choice of tree-growing parameters.

\begin{remark}
Tree-based methods, including DT's, generally cannot enforce continuity of the estimated response
$\hat{T}(t,\cdot)$. This might appear to be problematic given that the true timing value $T(t,\cdot)$ is continuous in $x$. However,
empirical studies, including previous analyses (see \cite{BouchardWarin10}), show that this is not a concern practically, especially for
estimating the zero-contour of $T(t,\cdot)$. In fact, piecewise constant fits appear preferable to piecewise linear (despite being more
``discontinuous'') since the latter are prone to finding spurious zero-contours.
An alternative would be to use interpolating methods, such as Gaussian processes \cite{PichenyGinsbourger13} that can respect continuity.
\end{remark}

Dynamic trees assume a Gaussian distribution for $Y|X$ in \eqref{eq:regress} and therefore the posterior predictive distribution
at location $x$ has a $t$-distribution (cf.~\cite[Sec 2.2]{GTP-trees11})
\begin{align}
Y(x) | \cZ^{(n)} \sim St \left( m^{(n)}_t(x), \hat{\sigma}^{2,(n)}_t(x), n_T -d -1 \right),
\end{align}
where $\hat{\sigma}^2_t(x)$ is the estimated predictive variance of the response at $x$. (Note that the above posterior uncertainty is for each tree-particle; the ultimate DT is a mixture of $M$ such estimators.) The Student-t
distribution has $n_T - d-1$ degrees of freedom, where $n_T$ is the number of samples in the leaf of the fitted tree $\mathcal{T}_n$ and $d$ is the dimension
of $\XX$.
 Using the Gaussian limit (i.e.~setting $n_T$ large) yields that the estimated timing value (which is the mean of $Y(x)$) satisfies
$$
T(t,x) | \cZ^{(n)} \sim \mathcal{N} \Bigl( \hat{T}^{(n)}(t,x), \frac{\hat{\sigma}^{2,(n)}_t(x)}{ n_T - d - 1} \Bigr).
$$
To evaluate the local loss functions in  \eqref{eq:EI-zc}-\eqref{eq:EI-sign} one can either use the above approximation averaging the parameters across the tree-particles population or directly consider the discrete posterior measure $F^{disc}_x$ in \eqref{eq:discrete-F} based on  the individual particles $m=1,\ldots,M$. In particular, the posterior mean is estimated as $\hat{T}(t,x) = \frac{1}{M} \sum_{m=1}^M \hat{T}_m(t,x)$, and the variance as $\hat{v}(t,x) = \frac{1}{M} \sum_{m=1}^M (\hat{T}_m(t,x) - \hat{T}(t,x))^2$. Similarly, one can easily evaluate the expected reduction in variance
\begin{align}\label{eq:DT-alc}
ALC_n(x) = \frac{\hat{\sigma}^{2,(n)}_t(x)}{ (n_T - d - 1)} - \frac{\hat{\sigma}_t^{2,(n+1)}(x)}{ (n_T-d)} \simeq \frac{\hat{\sigma}^{2,(n)}_t(x)}{ (n_T - d - 1)(n_T-d)},
\end{align}
where the latter expression is averaged over the tree-particles. The \texttt{dynaTree} package provides access to all these particle parameters (posterior mean, variance, degrees of freedom, etc.)

\subsection{Selecting New Sample Points} \label{sec:sampling}
In Section \ref{sec:al} we discussed construction of EI measures $EI_n(x)$ to guide the growing of the designs $\cZ^{(n)}$.
Naively, given the existing design $\cZ^{(n)}_t$, the ideal next site is then
$ x^{n+1,*}_t = \arg \sup_{x \in \cX} EI_n(t,x)$. Finding $x^{n+1,*}_t$ not only requires solving a $d$-dimensional optimization problem over $\cX$, but also puts full faith in the (myopic) EI heuristic. Consequently,
we employ probabilistic algorithms to approximate $x^{*,n}$. First, instead of optimizing over $\cX$, we randomly generate a finite candidate set $\cD$ of size $D$ and evaluate $EI_n(t,x'_\ell), x'_\ell \in \cD$ using \eqref{eq:ei-main}.
Second, we then use $\epsilon$-greedy sampling to pick $x^{n+1}_t$ from $\cD$, i.e.,
\begin{align}\label{eq:eps-greedy}
x^{n+1}_t = \left\{ \begin{aligned} \arg \min_{x' \in \cD} EI_n(t,x') & \text{ with prob } 1-\eps \\
x'_1 & \text{ with prob } \eps \end{aligned}\right.
\end{align}
The candidate set $\cD$ is generated using a Latin hypercube sample (LHS).
Allowing random choice of $x^{n+1}_t$ with probability $\epsilon  > 0$ ensures that the resulting design $\cZ^{(n)}$ becomes dense over entire $\cX$ as $n \to\infty$ which is sufficient to ensure global consistency of the estimated response.

\begin{remark}
The issue of converting EI measures into actual design site choices is common to all stochastic approximation settings. A typical solution is simulated annealing which corresponds to sending $\eps \equiv \eps(n) \to 0$ as $n \to \infty$ in \eqref{eq:eps-greedy}. See \cite{KushnerYinBook} and references therein for more details.
\end{remark}

\subsection{Adaptive Termination}
Availability of an approximation to the posterior error on the response $T(t,\cdot)$ in \eqref{eq:total-error} allows a fully adaptive scheme with performance guarantees. This is in contrast to basic LSMC, where the approximation architecture is fixed and design size $N$ must be specified a priori. With sequential design, it becomes possible to provide an
online indicator for when the overall grid $\{x^{1:n}_t\}$ is deemed sufficiently fine.
Indeed, for any fixed location $x$ and any $\eps > 0$, after sampling sufficient number of sample points, we have an empirical guarantee that $\PP_{0,x}( \sgn\hat{T}(t,X_t) = \sgn T(t,X_t) )> 1-\eps$ (ignoring the recursive nature of computing the timing values).

By integrating over $\cX$ one may combine the local error estimates into the global \eqref{eq:total-error}. Again, to avoid the associated high-dimensional integration, one simple termination criterion is the empirical average EI score of $\cZ^{(n)}$,
 $$ Err_t \simeq \frac{1}{D} \sum_{j =1}^{D} L_{n}(t,x_\ell) p(t,x_\ell | 0, X_0) \le Tol_t,$$
 where $Tol_t$ is the user-specified tolerance level for the $t$-th time step and $x_\ell \in \cD$ ranges over the picked candidate set $\cD$.
With the above considerations, one may implement RMC as an ongoing
statistical optimal design problem, perpetually proposing grid locations $x^n_t$ (across varying time steps) and incorporating them into the evolving estimates of $\hat{\mathfrak{S}}_t$. As time passes, the computed solution converges to the true one; in the meantime one
can always ``download'' the latest version of $\mathfrak{S}_{1:T}$. Thus, the algorithm can be set up in the background on a server,
updating whenever computing resources become available.

 \begin{remark}
By definition, there is very strong dependence between $T(t,\cdot)$ and $T(t+1,\cdot)$. Because these functions are
estimated backwards in time, a more accurate estimate of $T(t,\cdot)$ is needed for larger $t$'s. Based on \eqref{eq:egloff},
a simple rule is $Tol_t = C \cdot 3^{-t}$, which implies that the designs $\cZ_t$ should be of increasing size $N_t$ as $t$ grows. More precise analysis of the best allocation of $N_t$ across time-steps is beyond the scope of this paper and is left for future work.
\end{remark}

\subsection{Implementation Remarks}
New sequential design steps requiring updated fits over time adds overhead to the RMC scheme. We therefore discuss some of these computational budget concerns and our work-arounds which are summarized in Algorithm \ref{algo:Tree-OptStop} presenting our full implementation of sequential RMC.

While well-suited to sequential RMC framework, DTs are computationally intensive. This is a practical concern in the usual context where the total number of samples is in the thousands. We note however that it is sufficient to have a simple (i.e.~``rough'') fit $f^{(n)}$ during active learning, followed
by a final ``gold standard'' fit at the end. The intermediate fits are factually only used to compute the EI heuristics $EI_n(\cdot)$
and hence their main requirement is ease of update rather than precision. As such we have found it acceptable to reduce computation time by using a single particle $M=1$ dynamic tree, rejuvenated every few iterations. Once the full design $\cZ_t$ is generated, we then obtain the final estimate $\hat{T}(t,\cdot)$ (that will be used then
 used by Algorithm \ref{algo:forward-sim} for forward path simulation) using a large DT (eg.~$M=100$ particles).

Second, the sequential design step in Algorithm \eqref{algo:Tree-OptStop} requires initialization. We do this by the LSMC trick of generating $N_0$ paths $x^{1:N_0}_{1:T}$ starting from fixed $X_0$. This can also serve as a rough estimate of $p(t,x | 0,X_0)$ (e.g., for \eqref{eq:ei-main}) when the  density of $X_t | X_0$ is not available analytically.

Third, to speed up the procedure for the case where thousands of samples are needed, we consider batch schemes that simultaneously
 pick $N' \gg 1$ new design points without recomputing  $EI_n(\cdot)$. This is implemented by the additional loop in step 8 of Algorithm \ref{algo:Tree-OptStop} below, which is practically implemented as a multinomial sampling with replacement. To match the
randomization in \eqref{eq:eps-greedy}, we sample based on a potential
\begin{align}\label{eq:weights}
\PP( x^{n+1}_t = x'_\ell) \propto w_n(t,x) := \exp( -\beta_n (\bar{EI} \wedge EI_n(t, x'_\ell) ) ),
\end{align}
 where $EI_n$ are first normalized to have minimum of 0, and mean 1, so that the sampling weights $w_n$ are uniformly bounded.
 The parameter $\beta_n$ is similar to the cooling schedule in simulated annealing and is the analogue of $\eps$ in \eqref{eq:eps-greedy}.
Note that the design sites $\{x^n_t\}$ need not be distinct so there is no issue with picking the same site more than once during sampling. In examples below, $N'$ is in the range $N' \in [50,100]$.

\begin{algorithm}[ht!]
\caption{Tree-based Adaptive MC Algorithm for Optimal Stopping\label{algo:Tree-OptStop}}
{\fontsize{11}{11}\selectfont\begin{algorithmic}[1]
\REQUIRE $N_0$ -- number of initial grid points
\STATE $\mathfrak{S}_T \leftarrow \mathcal{X}$
\STATE Generate $N_0$ trajectories $x^{1:N_0}_{1:T}$ starting with $x^N_0 = X_0$ and using $x^n_{t+1} \sim p(t+1, \cdot | t, x^n_t)$
\FOR{$t=T-1,T-2,\ldots,0$}
\STATE $n \leftarrow N_0$
\STATE $\cZ^{(n)}_t \leftarrow \{x^{1:N_0}_t\}$
\STATE Using Algorithm \ref{algo:forward-sim} and $\hat{\mathfrak{S}}_{t+1:T}$ find $y^{1:N_0}_t$
\STATE Estimate the initial fit $f^{(n)}(\cdot)$
\WHILE{ the  current design needs refining}
\STATE Generate a fresh candidate set $\{x^{(n),j}\}$ $j=1,\ldots,D$ using LHS
\STATE Compute the expected improvement scores $EI_n(t,x^{(n),j})$ and weights $w_n(t,x^{(n),j})$ using \eqref{eq:weights}
\FOR{$n'=1,\ldots,N'$}
\STATE Sample (with replacement) according to weights $w_n(t,\cdot)$ a new design point $x^{n+n'}_t$
\STATE Simulate forward trajectory $x^{n+n'}_{t+1:T}$
\STATE Using Algorithm \ref{algo:forward-sim} and $\hat{\mathfrak{S}}_{t+1:T}$ find $y^{n+n'}_t$
\STATE Update the design $\cZ^{(n+n')}_t \leftarrow \cZ_t^{(k)} \cup (x^{n+n'}_t,y^{n+n'}_t)$
\ENDFOR
\STATE $n \leftarrow n+N'$
\STATE Update the fit $f^{(n)}$ based on the latest $\cZ^{(n)}_t$
\ENDWHILE{// now $n=N_t$}
\STATE Generate final estimate $\hat{T}(t,\cdot)$ based on the ultimate $\cZ^{(N_t)}_t$ at time step $t$ and corresponding
$\hat{\mathfrak{S}}_{t}$
\ENDFOR
\STATE Simulate forward trajectories $\tilde{X}^{1:N}_{0:T}$ from $\tilde{X}^{n}_0 = X_0$ using $\hat{\mathfrak{S}}_{0:T}$
\RETURN $\hat{V}^{(N)}(0,X_0) \simeq \frac{1}{N} \sum_{n=1}^N h_{\tau^n}(\tilde{X}^{n}_{\tau^{n}})$ as estimate of $V(0,X_0)$
\RETURN Estimated policy $\{\hat{\mathfrak{S}}_{0:T}\}$.
\end{algorithmic}}
\end{algorithm}

\begin{remark}
Observe that Algorithm \ref{algo:Tree-OptStop} only uses estimates of $C(t,x)$  or $V(t,x)$ in the very last step. Nor does it require access to $T(t,x)$ since all the key objects are described only in terms of $\hat{\mathfrak{S}}_t$. Thus, one could dispense with the regression step altogether, replacing it with a classification step instead, the aim being to classify input space $\cX$ into $\hat{\mathfrak{S}}_t$ and its complement. Such classification MC was proposed in \cite{Picazo02}. However, in the context of Bermudan option pricing the very high noise and skew of $Y(x)$ cause significant numerical challenges. In particular, $\PP( \sgn Y(x) = \sgn T(t,x) ) < 0.5$ is possible, i.e.~the skew in the sampled payoffs $Y(x)$ can be so extreme that its sign is usually the opposite of the sign of the actual timing value (this happens for shallow in-the-money American Put where the majority of paths will indicate that immediate exercise is preferable).
\end{remark}

\subsection{Algorithm Complexity}
The total number of simulations in Algorithms \ref{algo:mc-generic}-\ref{algo:Tree-OptStop} is
\begin{align}\label{eq:num-of-sims}
TOTSIM := \sum_{t=1}^T N_t (\tau_t-t)
\end{align}
since each of the $N_t$ $\XX$-paths at step $t$ is simulated from $t$ until the corresponding stopping time $\tau_t$. Note that $TOTSIM$ is stochastic and strongly affected by the chosen designs $\cZ_t$. At first glance,  it looks like $\E[TOTSIM] = \OO( T^2)$ since $\tau_t  t\simeq T -t$. However, because most of the design points are by construction close to $\partial \mathfrak{S}_t$, we observed in our experiments that the resulting $\tau_t$ is typically much smaller than $T$ so that $\E[\tau_t - t]$ is almost independent of $t$.

\begin{remark}\label{rem:augment}
We also note that \eqref{eq:num-of-sims} assumes fresh re-generation of paths $x^{(t),1:N_t}_{t:T}$ at each step. Algorithmically these forward simulations can be put to use by
\emph{augmenting} existing designs $\cX_s$ for $s>t$ with $x^{(t),1:N_t}_{s}$ and accordingly recomputing $\hat{T}(s,\cdot)$. This
leads to a looping adaptive backward-forward estimation that is potentially very powerful.
\end{remark}

A completely different way to reduce simulation effort is
to implement a blend of value and policy iteration. Recall that the RMC
framework is a pure policy iteration. At the other extreme, one may consider pure value iteration as in \eqref{eq:recurse-V}. Recursive solution of \eqref{eq:recurse-V} using Monte Carlo requires only simulation of the one-step paths
$(X_{t:t+1})$. Thus, the conditional variance is roughly proportional to $\tilde{\sigma}^2(t,x) \propto \mathbb{V}ar(X_{t+1}|X_t=x)$
which is much less than $\sigma^2(t,x) = \mathbb{V}ar(X_{\tau} | X_t)$ experienced during policy iteration. The resulting method can
reduce $\sigma^2(t,x)$ by an order of magnitude. However, it also introduces a much stronger dependence between $\hat{V}(t,\cdot)$ and
$\hat{V}(t+1,\cdot)$ since errors in estimating $V(t+1,\cdot)$ \emph{necessarily} feed into the estimate of $V(t,\cdot)$ as well.
Moreover, value iteration demands solving a full regression problem (rather than contour-finding) at each step. Nevertheless, this could still be computationally advantageous if one can dramatically cut down the number of samples needed to learn $V(t,x)$.

\section{Numerical Examples}\label{sec:numeric}
To illustrate the proposed algorithm we benchmark our results using the examples in the recent paper of Bouchard and
Warin~\cite{BouchardWarin10}. This set of examples considers pricing of Bermudan options on several assets for a variety of payoff
types.
Slightly abusing notation, we henceforth use $t$ to denote financial physical time. The exercise rights are discretized using $\Delta T$, so that RMC operates on the discrete grid $\tau \in \{ 0,\Delta t, \ldots, T\}$. Let
\begin{align}\label{eq:gbm}
S^{(j)}_t = S_0^{(j)} \exp \left({ (r-\frac{1}{2} \sigma^2) t + \sigma  W^{(j)}_t}\right), \qquad j=1,\ldots, d,
\end{align}
where $W^{(j)}$ are $d$ independent Wiener processes. Thus, $S^{(j)}$ are i.i.d.~log normal and \eqref{eq:gbm} corresponds to a multi-dimensional Black-Scholes model where all the volatilities $\sigma$ are equal.
We denote by $\XX \equiv (S^{(1)}, \ldots, S^{(d)})$ the vector of $d$ asset prices to be
considered.  We then study payoffs of the form
\begin{align*}
h^{Put}_t( X_t) &= e^{-r t}\left(K - \prod_{j=1}^d S^{(j)}_t \right)_+, &   h^{Basket}_t( X_t) &=  e^{-r t}\left(K - \frac{1}{d} \sum_{j=1}^d S^{(j)}_t
\right)_+ 
\end{align*}
where $K \in \R_+$ is the strike. These options correspond
to the geometric and arithmetic basket Puts. In particular, either payoff with $d=1$ corresponds to
the classical Bermudan Put in the Black-Scholes model. Using properties of the log-normal distribution and independence, we have that the product
$\check{S}_t := \prod_{j=1}^d S^{(j)}$ is again log-normal. Therefore, pricing of the contract with payoff $h_{Put}(\XX)$ can be reduced to pricing $h_t(\check{S}) = e^{-r t}(K - \check{S})_+$ within a one-dimensional setting which  can be done via efficient and highly accurate 1-d analytic methods.
On the contrary, the basket Put payoff $h^{Basket}(\cdot)$ does not admit any dimension
reduction and has been used in the literature to benchmark multi-dimensional American payoffs.

\begin{table}[ht!]
\begin{center}
\begin{tabular}{lc|lc} \hline
Initial condition & $S^{(1)}_0=40$ & Interest rate & $r=0.06$ \\
 & $S^{(2)}_0 = 40$  & Discretization & $\Delta t =0.04$ \\
Volatility & $\sigma = 0.2$ & Strike & $K=40$ \\ Horizon & $T=1$ & Payoff & $h^{Basket}(\cdot)$ \\
 \hline
\end{tabular}
\end{center}
\caption{Parameters for Section \ref{sec:example-1d} \label{tbl:params}}
\end{table}

To highlight the impact of sequential design we  kept the rest of the Algorithm \ref{algo:Tree-OptStop} at its most basic, in particular using the same design size $N_t$ across all time steps, and doing full path regeneration at each step (cf.~Remark \ref{rem:augment}). We expect that significant further savings can be extracted by optimizing the budget across backward time-stepping.

\subsection{Low-Dimensional Examples}\label{sec:example-1d}
We begin by revisiting the classical situation of a one-dimensional American Put option where $d=1$, $h_t(S) = e^{-r
t}(K-S)_+$ and the rest of the parameters are from \cite{LS}, see Table \ref{tbl:params}. In this case $V(0,S_0) = 2.314$. In one
dimension, the stopping region for the American Put is an interval $[0,\underline{s}(t)]$ so the contour-finding problem reduces to
finding the unique value $\underline{s}(t)$ for each time-step $t$. We similarly consider the two-dimensional basket Put version of the above $X_t \equiv ( S^{(1)}_t, S^{(2)}_t)$, using i.i.d.~$S^{(1)}$
and $S^{(2)}$ with payoff $h_t(x) = e^{-r t}( K - (S^{(1)}_t + S^{(2)}_t)/2)_+$. In that case, the stopping region is a connected
bounded subset of $\R^2_+$ containing the origin, i.e., $\mathfrak{S}_t := \{ (s_1, s_2) : s_1 \le \underline{s}_t(s_2) \}$.

\begin{figure}[ht!]
\centering 
\includegraphics[scale=0.8]{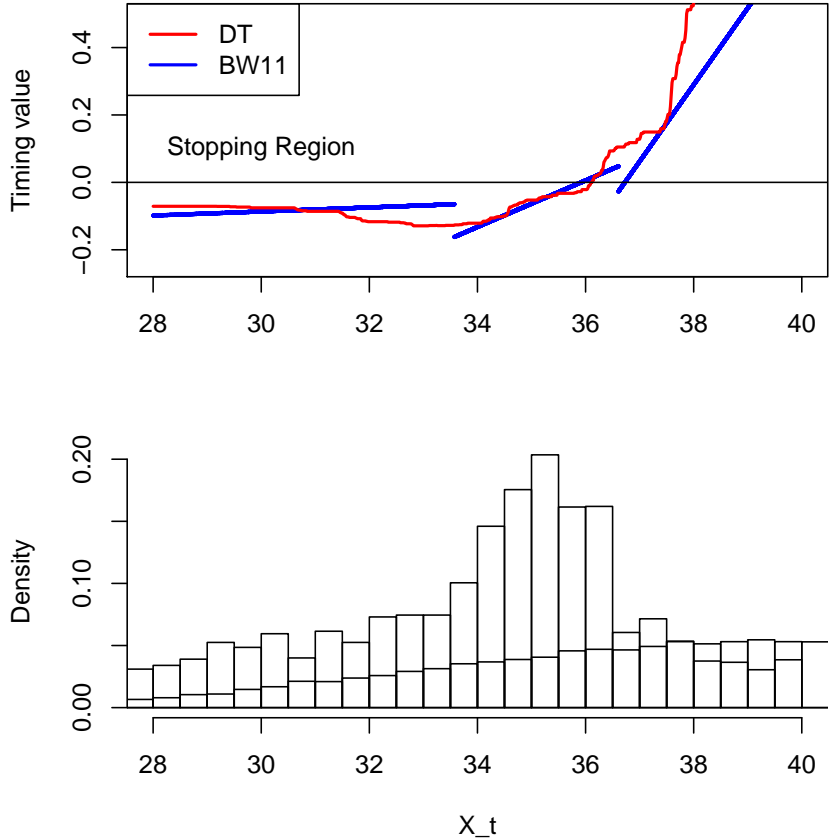}
\begin{minipage}{0.92\textwidth}
\caption{Comparison of the Bouchard-Warin (BW11) \cite{BouchardWarin10} and dynamic trees (DT) fits for the example of Section
\ref{sec:example-1d}. The top panel shows the estimated $T(t,x)$ for $t=0.8$. The bottom panel shows the distribution of the respective
grids $\{x^n_t\}$. BW11 used a design of $40,000$ samples grouped into $N_p=8$ piecewise-linear fits; DT used a design of $5,000$ with constant
leafs. \label{fig:compBW-1d}}
\end{minipage}
\end{figure}

Figure \ref{fig:compBW-1d} shows an example of using Algorithm \ref{algo:Tree-OptStop} to estimate $T(t,x)$ for the 1-d case at $t=0.8$. We generated a design with $N_t=5000$ locations using  the $EI^{ZC}$ heuristic
 from \eqref{eq:ei-main} (see the already discussed Figure \ref{fig:active-learning}). We used a single-tree $M=1$ constant-leaf DT that was rejuvenated after every 500 sample points during active learning, and $M=10$ constant-leaf trees for the final fit of $\hat{T}(t,\cdot)$. To augment the designs $\cZ^{(n)}$ we started with $N_0=1000$ points sampled from $p(t,\cdot | 0, X_0)$ and ran 40 iterations of the \texttt{while} loop in Algorithm \ref{algo:Tree-OptStop}, adding $N'=100$ points in batch per iteration. Thus, the ultimate size of the grids $N_t \equiv 5,000$ was predetermined and constant over time-steps $t$. During each iteration we proposed an LHS set of $D=500$ locations, and sampled from these locations using the potential in \eqref{eq:weights} with $\beta=0.5$. Based on our experiments the DT implementation had comparable performance over a wide range of the above parameters.

We then compared our DT-based implementation against the Bouchard-Warin (BW11) \cite{BouchardWarin10} algorithm, which was used as a benchmark of a non-adaptive grid RMC with an advanced regression architecture. As can be seen in the Figures \ref{fig:active-learning}-\ref{fig:compBW-1d}, sequential design  allows a dramatic refocusing of the design $\cZ_t$. Instead of designs centered around the
unconditional mean $\E[ X_t |X_0] \sim 40$, the dynamic regression approach focuses on the location of the zero-contour. This reduces
the number of samples required by an order of magnitude without sacrificing precision of the fit. Moreover, we observe that even
constant-leaf trees can, in \emph{ensemble}, provide a robust fit to the nonlinear shape of the contour. The observed
artifacts of the DT-based $\hat{T}(t,x)$ at the edges of the plots (in particular severe underestimate of the true $T(t,x)$ on the extreme right $x \ge 38$) are to be expected given the low density of points there, but are
practically irrelevant since these regions are almost never hit by the paths $(X_t)$.

\begin{figure}[ht!]
\begin{tabular}{cc}
\begin{minipage}{0.475\textwidth}
\ \ \includegraphics[width=0.95\textwidth,trim=0.05in 0.15in 0.15in 0.15in,clip=true]{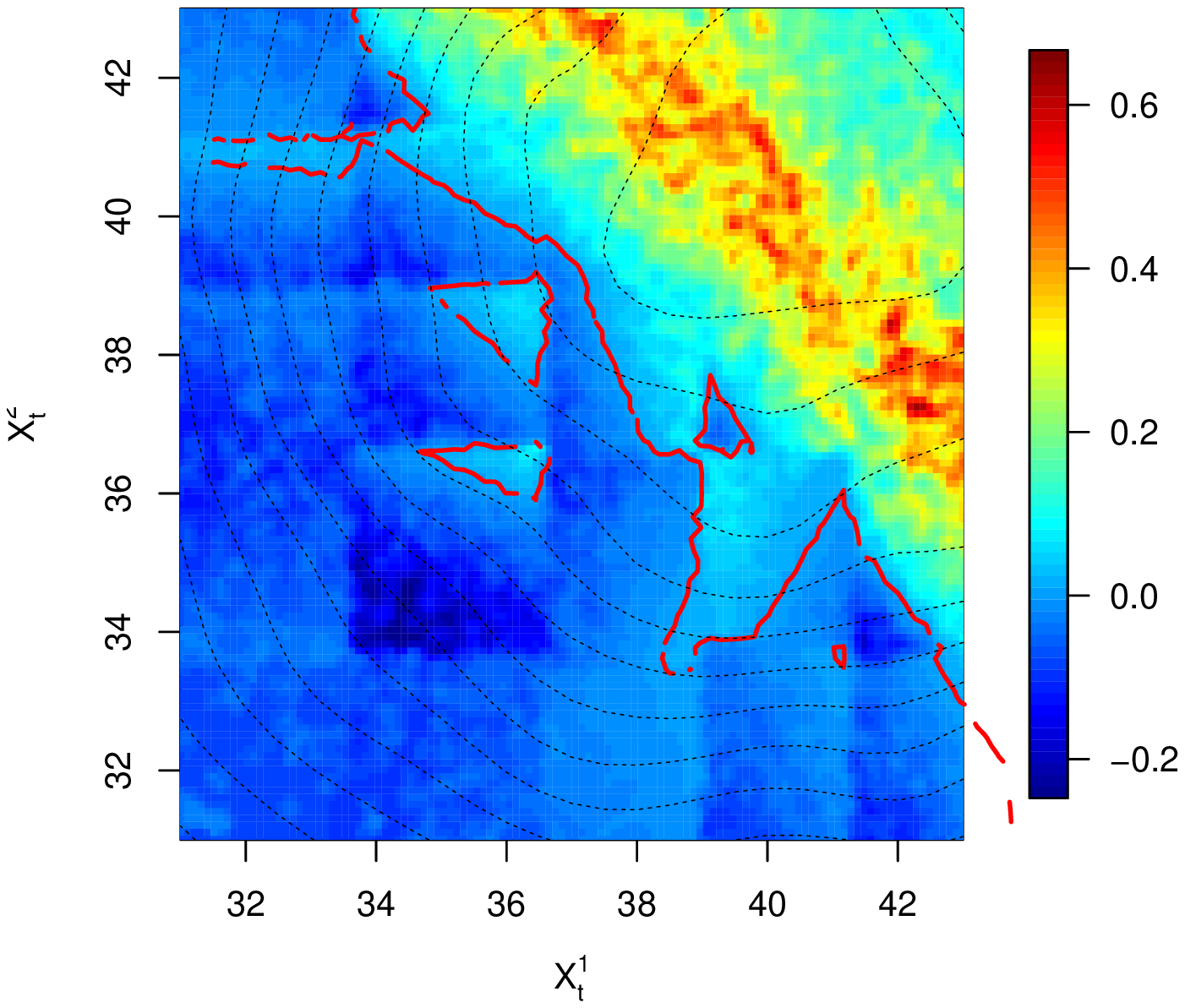}
\end{minipage} &
\hfill
\begin{minipage}{0.475\textwidth}
\includegraphics[width=0.95\textwidth,trim=0.05in 0.15in 0.15in 0.15in,clip=true]{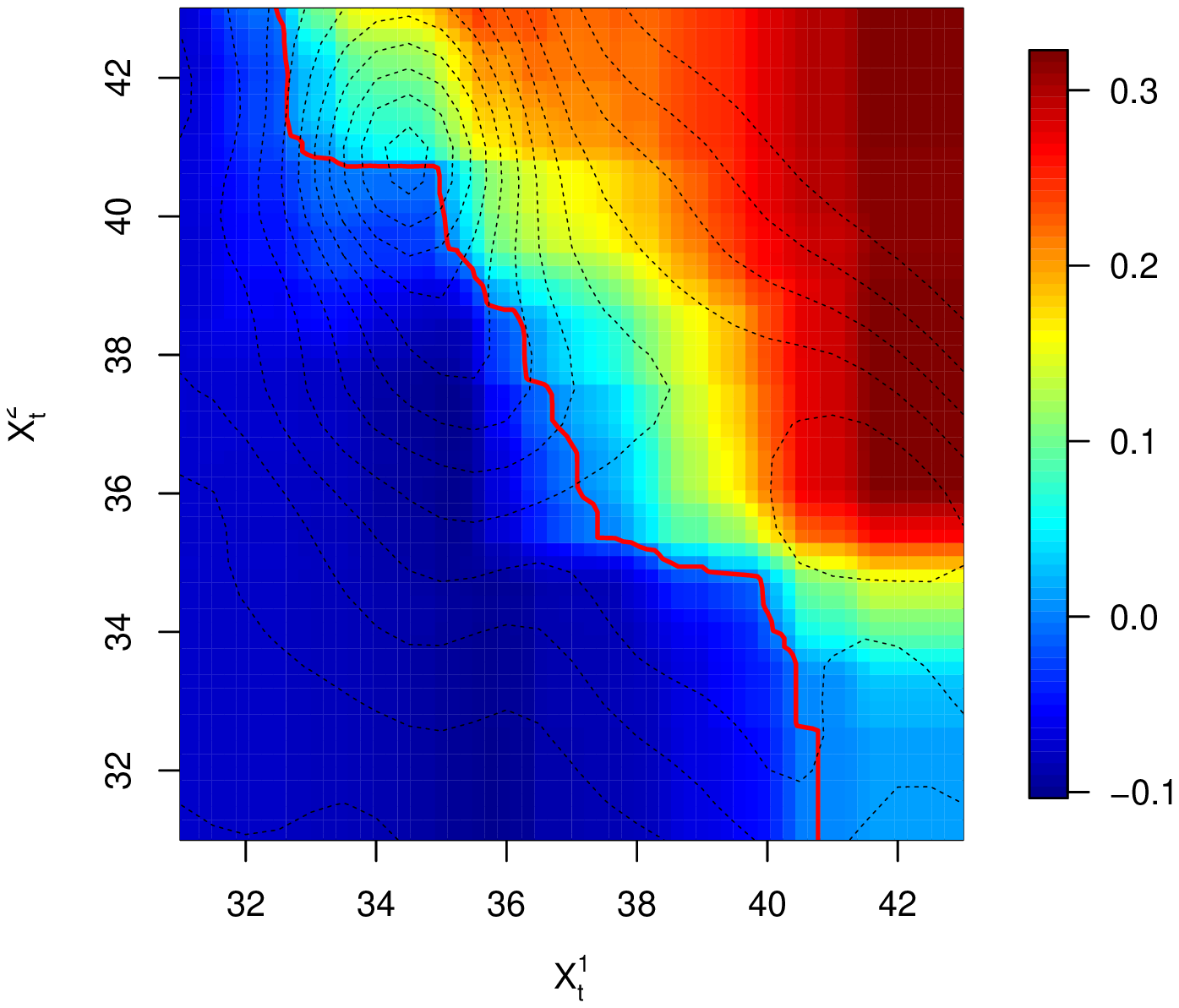}
\end{minipage} \\
 \small{BW11} & \small{DT}  \end{tabular}
\begin{minipage}{0.92\textwidth}
\caption{Comparison of the Bouchard-Warin (left) \cite{BouchardWarin10} and dynamic trees (right) fits in 2 dimensions. The heatmap
indicates the levels of the estimated timing value $T(t,x)$; the corresponding zero-contour $\{ x: T(t,x) = 0 \}$ is highlighted in red.
The other contours show the kernel density estimate of the distribution of the respective grids $\{x^n_t\}$. BW11 used $40,000$ samples
grouped into $(N_p)^d=8^2$ piecewise-linear fits; DT used $5000$ total samples with constant leafs. \label{fig:compBW-2d}}
\end{minipage}
\end{figure}

Similar phenomena are demonstrated in Figure \ref{fig:compBW-2d} in a 2-d setting where the zero-contour is a smooth curve. The plot compares a DT implementation with $N_t = 5000$ against a BW11 implementation with $N_t = 40,000$. Again, sequential design recenters the design sites towards in-the-money and is able to adaptively generate a much more refined estimate of $\partial \mathfrak{S}_t$. Since the scheme of BW11 relies on piecewise linear fits, even with a large design size the respective stopping region is not connected (one clearly sees the boundaries of the cells in $\cX$ used for the local regressions as discontinuities in the estimate of $T(t,x)$). This pitfall is common to any ``global'' statistical modeling of $T(t,\cdot)$ and would be even more severe with the usual LSMC use of global basis functions $B_r(x)$. In contrast to these jagged contours, the hierarchical local-based tree fit organically generates much ``smoother'' estimates of the stopping boundary. The overall estimate of the value function $\hat{V}(0,X_0)$ produced by the two methods is comparable despite our method using a design 8 times smaller.
While BW11 had a total of $N \cdot T =1.25 \cdot 10^6$ simulation steps, the dynamic regression algorithm used $TOTSIM = 8.39 \cdot 10^5$ total simulations, a $1/3$ reduction.

\begin{figure}[ht!]
\begin{center}
\includegraphics[scale=0.63]{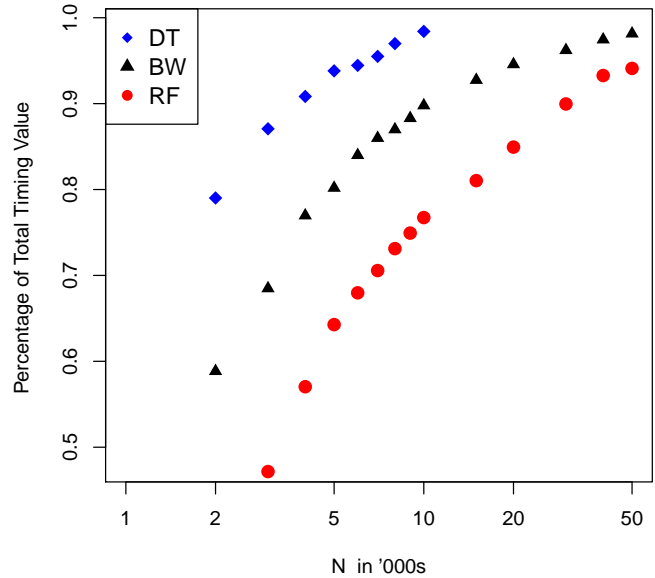}
\begin{minipage}{0.92\textwidth}
\caption{Performance of (i) a dynamic tree sequential RMC (DT), (ii) localized linear regression (BW) LSMC \cite{BouchardWarin10} and
(iii) random forest LSMC (RF) methods as a function of design size $N$ for a two-dimensional basket Put. Each reported
value is an average of $\hat{V}(0,X_0)$ across 10 runs of the RMC. \label{fig:rel-error}}
\end{minipage}
\end{center}
\end{figure}

From a different angle, Figure \ref{fig:rel-error} compares the performance of three
different RMC methods on this two-dimensional example in terms of the design size $N \equiv N_t$.  Namely, we plot the results of a DT
sequential RMC against LSMC implementations using (i) a local regression (BW11) and (ii) a random forest (RF). We provide the RF LSMC as another example of a hierarchical non-parametric regression---comparing RF LSMC and DT RMC therefore isolates the impact of adaptive design. Besides the regression tool used, RF and BW11 LSMC implementations were identical.  To control for the Monte Carlo error, we fixed a common set of out-of-sample paths $\tilde{X}^{1:{N}}_{0:T}$, with ${N} = 50,000$ which were used to produce the final estimate $\hat{V}^{(N)}(0,X_0)$.
Also, to get a better sense of the relative errors being made by RMC in terms of practical accuracy, we plot the percentage of total
timing value (extrinsic value) estimated by the RMC scheme, $\frac{ \hat{V}(0,X_0)- \underline{v}(0,X_0)}{ V(0,X_0) -
\underline{v}(0,X_0) }$, relative to the benchmark value of $V(0,X_0) = 1.464$  and the intrinsic value $\underline{v}(0,X_0) =
\E_{0,X_0}[ h_T(X_T)] = 1.230$.

As can be observed, adaptive placement of the design points $\{ x^n_t\}$ leads to savings of 80-90\%, i.e., a non-adaptive design of $N_t=25,000$ is comparable to an adaptive design of $N_t=3000$ points. We also see that the RF fits underperform compared to piecewise linear fits of BW11. This could be
because the stopping boundary is effectively a hyperplane in terms of the average
$(S^{(1)}_t + S^{(2)}_t)/2$ and is poorly approximated by constant-leaf trees generated by RF regression.

Finally, in Table \ref{tbl:6d} we show the performance of the methods on a 6-dimensional example given in \cite{BouchardWarin10}. With
higher dimensions, more simulations are needed to approximate the hypersurface defining the stopping boundaries $\partial
\mathfrak{S}_t$. The dynamic regression approach continues to require far fewer paths for same level of accuracy compared to standard
LSMC. The relative performance gain varies depending on the payoff type considered. For the geometric Put payoff $h_{Put}(\cdot)$, DT
only slightly beats BW11 algorithm in terms of simulation effort to achieve same precision. However, note that this example is
degenerate, the most important summary statistic being the one-dimensional product $\check{S}_t$. As a result, the stopping boundary is
``almost'' a threshold rule (i.e.~a hyperplane) in $\check{S}_t$. This low-dimensional structure plus the high nonlinearity
corresponding to $\check{S}_t$ makes linear projections onto $span( S^{(1)}_t, \ldots, S^{(6)}_t)$ that are performed by all the
considered regression methods rather poor. A simple solution (which we do not pursue here to avoid digressions) would be to include
$\check{S}_t$ as one of the state-variables for the statistical model of $\hat{T}(t,\cdot)$, similar to the strategy commonly employed in classical LSMC. For the
basket Put payoff $h^{Basket}(\cdot)$, we find that dynamic regression can save up to 50-60\% of simulation effort. In both cases,
running times of DT RMC remain longer since the regression overhead still dominates.

\begin{table}[ht!]
$$ \begin{array}{ccccc} \hline
 & \multicolumn{4}{c}{\textbf{Basket Put}} \\ \hline \hline
 \text{(in 000's)} & N = 6.4  & N=9.6  & N = 64 & N = 128 \\ \hline
BW &  0.0174 & 0.0176 & 0.0181 & 0.0181  \\ TOTSIM \text{('000s)} & 76.8 & 115.2 & 768 & 1536 \\ DT & 0.0181 & 0.0182 & - & - \\
TOTSIM \text{('000s)} & 305 & 435 & - & - \\ \hline
 & \multicolumn{4}{c}{\textbf{Geometric Put}} \\ \hline \hline
\text{(in 000's)} & N = 6.4 & N=9.6 & N = 64 & N = 128\\ \hline BW & 0.1054 & 0.1069  & 0.1095 & 0.1098 \\
TOTSIM \text{('000s)} &
76.8 & 115.2 & 768 & 1536 \\ DT & 0.1092 & 0.1095 & - & - \\
TOTSIM \text{('000s)} & 287 & 412 & - & - \\ \hline
\end{array}$$
\begin{minipage}{0.92\textwidth}
\caption{Performance of different RMC algorithms for a 6-d Bermudan Put. Throughout, $T=1,r=0.05,\Delta t=1/12,K=1$ and the asset prices
are Geometric Brownian motions with $S^j(0)=1$ and volatility $\sigma=0.2$. Dynamic trees (DT) used constant leafs with $M=5$ for intermediate
fits and linear leafs with $M=4$ particles for final fits. Each reported value is an average of $\hat{V}(0,X_0)$ across 10 runs of RMC.
\label{tbl:6d}}
\end{minipage}
\end{table}

\subsection{Stochastic Volatility Models}\label{sec:stoch-vol}
As our second application we consider pricing of Bermudan options within stochastic volatility models. These examples are more computationally challenging since (i) simulating $\XX$ is more costly and (ii) the structure of the optimal stopping set is more complex.

We suppose that under the pricing measure the state process $X = (S,Y)$ follows
\begin{align}\label{eq:stoch-vol}
dS_t = r S_t \,dt + \exp(Y_t) \, dW^{(1)}_t \\
dY_t = \delta (m - Y_t) \,dt + \nu \, dW^{(2)}_t
\end{align}
where the standard Brownian motions $(W^{(1)}, W^{(2)})$ have correlation $d \langle W^{(1)}, W^{(2)} \rangle_t = \rho dt$. Thus, the log-volatility factor $(Y_t)$ is mean-reverting to the level $m$ and has vol-vol of $\nu$. While $(Y_t)$ is an autonomous Ornstein-Uhlenbeck process with analytic Gaussian density, there are no explicit formulas for the transition density of the pair $(S_t,Y_t)$. Consequently, a numerical discretization scheme must be used to simulate the corresponding paths of $(S_t)$; we apply a basic Euler discretization with step $\delta t = 0.1/252$ (i.e.~0.1 days). Also,
to recover the density $p(t,\cdot|0,X_0)$ we use a kernel density estimator based on the locations $x^{1:N_0}_t$.

Following \cite{Rambharat11}, we consider pricing a Bermudan Put on $(S_t)$ with payoff $h(X) \equiv h(S,Y) = (K - S)_+$ and three parameter sets specified in Table \ref{tbl:SV-params}.
The horizon $T$ is in daily units (i.e.~$T=10$ means 10 days or 10/252 years), and exercise is assumed to be daily.
All three examples start significantly in-the-money, which is the situation most favorable to standard RMC since few paths of $X_{0:T}$ ever go out-of-the-money where they are wasted. As shown in Table \ref{tbl:SV-results} our adaptive approach outperforms the  benchmark by more than an order of magnitude across the full  spectrum of cases considered. The first Set I of parameters is reasonably realistic and at this time horizon makes the stopping boundary almost constant in $S_t$. Hence, it structurally resembles a 1-d Bermudan Put. The second set of parameters has a longer horizon and very high vol-vol $\nu$, as well as almost no mean-reversion. In that case, the boundary is truly two-dimensional. Finally, set III has even more extreme vol-vol which makes the Put almost European, i.e.~the probability of exercise is very small and $T(t,\cdot)$ is positive nearly everywhere. The last two cases are challenging for a simulation-based method since they put $\partial \mathfrak{S}$ at the ``edge'' of $\mathcal{X}$ and hence hard to identify without focused sampling there.

\begin{table}[ht!]
\begin{center}
\begin{tabular}{llcc} \hline
 & SDE params $(\delta, m, \nu, \rho)$ & Put params $(T,\Delta t, r,K, X_0)$ & DT-RMC params, $(D,\beta,N_0)$ \\ \hline \hline
Set I &   $(3.3, -0.583,  0.5, -0.055)$ & $(10,1,0.055,23,(20, \log(0.5)) )$ & (500,0.5,500)\\
Set II &  $(0.015,2.950, 3, -0.03    )$ & $(50,1, 0.0225,100, (90, \log(0.35)) )$ & (500,0.5,500) \\
Set III &  $(0.025,1.314, 4.5, -0.05)$ &  $(25, 1, 0.025,19, (17, \log(0.35) ) )$ & (500,0.5,500) \\\hline
\end{tabular}
\caption{Parameters for Section \ref{sec:stoch-vol}. All SDE/option parameters are from \cite{Rambharat11} corresponding to their Set I=Experiment 1, Set II=Experiment 5, Set III=Experiment 9. The DT-RMC parameters correspond to the size of candidate set $D$, the exponent $\beta_n$ in \eqref{eq:weights}, and the initial design size $N_0$. }  \label{tbl:SV-params}
\end{center}

\end{table}

\begin{table}
$$\begin{array}{l|ccc} \hline \\
\text{Method} & \text{Set I} & \text{Set II} & \text{Set III} \\ \hline \hline
\text{BW11 } N_t=32000 &  3.038 & 16.26 &  2.86\\
\text{BW11 } N_t=80000 &  3.044 &  16.42 &  2.89 \\
\text{DT  } N_t=3000 &  3.043  & 16.51 &  2.93 \\
\text{DT  } N_t=5000 &  3.045 &  16.60 &    2.94\\ \hline
\end{array}$$
\caption{The BW11 \cite{BouchardWarin10} method used $5^2$ partitions and linear fits in each partition. The DT method used fixed-size designs with batches of $N'=50$ for $N_t = 3000$ and $N'=90$ for $N_t=5000$, and $M=4$ constant-leaf trees throughout. All other parameters are in Table \ref{tbl:SV-params}.}
\label{tbl:SV-results}
\end{table}

\section{Discussion}\label{sec:discuss}
The presented sequential RMC approach makes several interrelated innovations compared to the existing paradigm. First, we reformulate the core step of LSMC as a contour-finding problem with the associated loss function \eqref{eq:loss}. While this view has been known before
\cite{Egloff05,Belomestny11}, to our knowledge it has never been computationally exploited. As we show, it makes a dramatic difference in terms of appropriate regression methods to apply \emph{empirically}. Reformulating the regression sub-problem as finding the contour of a noisily observed response function allows application of frameworks from machine learning, stochastic
optimization and computer experiment design.
Second, we develop and make full use of the path-regeneration feature of Algorithm \ref{algo:mc-generic} which allows adaptive grid placement. Hitherto, path-regeneration was only a theoretical device to facilitate convergence proofs; in our setup it is a crucial piece of the algorithm. Adaptive grids combined with adaptive regressions fully exploit the localized nature of the loss function. Moreover, they allow grid sizes to be refined over time to control the backward error propagation.

Third, we introduce the framework of sequential design, which has already been very successful in stochastic optimization and meta-modeling literatures, to the context of RMC. Sequential design jointly optimizes the stochastic grids and the estimated fits to wring maximum computational efficiency. As we show these gains are up to an order of magnitude. Fourth, we view the regression steps of RMC through the lens of response surface modeling, focusing on the full posterior distribution of the unknown timing value instead of the common point estimate. This new perspective facilitates evaluation of the empirical loss function \eqref{eq:empirical-error} and therefore provides novel, online estimates of the regression error. Here we propose to use dynamic trees to approximate \eqref{eq:empirical-error}, but other choices are also possible, such the local
Gaussian processes explored in \cite{GL14stat}.

Sequential RMC is a general-purpose algorithm that can apply to any discrete-time optimal stopping problem. Moreover, our ideas are not limited to optimal stopping, and could be ported to optimal switching, impulse control and other
settings. Indeed, related approaches in more general DP problems can be traced back to  \cite{ChenRuppert99}, fifteen years ago. A particularly relevant extension would be to optimal switching problems where the single stopping decision is replaced by a sequence of actions $(\tau_k, \xi_k)$ where $\xi_k$ is the $k$-th action taken at stopping time $\tau_k$, $\tau_1 < \tau_2 \ldots < T$. Usually, the action space $\mathcal{A} \ni \xi_k$ is discrete and very small (e.g.~trivial $|\mathcal{A}|=1$ or $|\mathcal{A}|=2$ for gas storage). Sequential RMC for optimal switching requires simultaneous modeling of $\ell$ switching sets $\mathfrak{S}^\ell_t$ introducing an extra dimension to the sequential design sub-problem.

\subsection{Computational Improvements}
In view of the above discussion, the presented implementation in Algorithm \ref{algo:Tree-OptStop} can be further enhanced in several directions. To begin, empirical loss estimation permits usage of adaptive termination criteria for the overall RMC algorithm, as well as allocation of computational budget across time-steps. This requires more analysis of the dependence between design size, empirical
error and error back-propagation in time.

Next, as mentioned before, the nature of the loss function suggests that one may implement a classification model that directly focuses on finding the classification boundary in lieu of searching for the zero-contour within a regression model. While a direct classification methodology does not work well \cite{Picazo02}, there would seem to be potential for a sequential design strategy based on a full posterior classification vector; a full discussion of the corresponding algorithm is beyond the scope of this paper and will be presented in future work.

Finally, while sequential design has been a very active research area for the past 20 years, the RMC context presents several particular challenges. In particular, RMC distinguished by its collection of dependent classification problems indexed by $t$ and the presence of very low signal-to-noise ratio. Resolving these features allows use of sequential methods borrowing ideas from meta-modeling of \emph{deterministic} experiments, and is also separately investigated in \cite{GL14stat}.

\bibliographystyle{acmtrans-ims}
\bibliography{../masterBib}

\end{document}